\documentclass[a4paper,11pt]{article}
\usepackage{jheppub}
\usepackage{dsfont}
\usepackage{amsmath,amssymb,amscd,amsfonts,mathtools}

\DeclareMathOperator{\Sinh}{Sinh}

\DeclareMathOperator{\Tanh}{Tanh}

\usepackage[toc,page]{appendix}
\usepackage{epsfig}
\usepackage{epstopdf}
\usepackage{latexsym}
\usepackage{graphicx}
\usepackage{placeins}
\usepackage{floatrow}
\usepackage{subfig}
\usepackage{caption}
\usepackage{booktabs}
\usepackage{bbm}
\usepackage{color}
\usepackage{physics}
\usepackage{stackrel}
\usepackage{tensor}
\usepackage{tikz}
\usetikzlibrary{matrix}
\usetikzlibrary{decorations.markings,calc,shapes,decorations.pathmorphing,patterns,decorations.pathreplacing,arrows.meta}
\usetikzlibrary{positioning}
\usepackage{hyperref}

\makeatletter
\newcommand\HUGE{\@setfontsize\Huge{50}{60}}
\makeatother 

\pdfoutput=1
\makeatletter
\def\@fpheader{\relax}
\makeatother

\usepackage{wasysym}

\def\kket#1{{|#1\rangle\!\rangle}}

\renewcommand{\ket}[1]{|#1\rangle}
\renewcommand{\bra}[1]{\langle#1|}

\DeclareMathSizes{10}{10}{7}{7}

\title{Transport across interfaces in symmetric orbifolds}

\author{Saba Asif Baig,}
\author{Sanjit Shashi}
\affiliation{Theory Group, Weinberg Institute, Department of Physics, University of Texas,\\
\phantom{}\hspace{0.5cm} 2515 Speedway, Austin, Texas 78712, USA.}
\emailAdd{sbaig@utexas.edu}
\emailAdd{sshashi@utexas.edu}

\abstract{We examine how conformal boundaries encode energy transport coefficients---namely transmission and reflection probabilities---of corresponding conformal interfaces in symmetric orbifold theories. These constitute a large class of irrational theories and are closely related to holographic setups. Our central goal is to compare such coefficients at the orbifold point (a field theory calculation) against their values when the orbifold is highly deformed (a gravity calculation)---an approach akin to past AdS/CFT-guided comparisons of physical quantities at strong versus weak coupling. At the orbifold point, we find that the (weighted-average) transport coefficients are simply averages of coefficients in the underlying seed theory. We then focus on the symmetric orbifold of the $\mathbb{T}^4$ sigma model interface CFT dual to type IIB supergravity on the 3d Janus solution. We compare the holographic transmission coefficient, which was found by \cite{Bachas:2022etu}, to that of the orbifold point. We find that the profile of the transmission coefficient substantially increases with the coupling, in contrast to boundary entropy. We also present some related ideas about twisted-sector data encoded by boundary states.}

\begin{document}	
\maketitle
\flushbottom
\vfill
\pagebreak

\section{Introduction}

2-dimensional conformal field theories with boundaries have a long history in the literature \cite{Cardy:1984bb,Cardy:2004hm}, with applications to both condensed matter theory \cite{Affleck:1995ge,Affleck2000} and worldsheet string theory \cite{Sagnotti:1987tw,Polchinski:1995mt,Gaberdiel:2002my,Recknagel:2013uja}. However, given a generic (particularly, irrational) theory, the underlying classification principles of such boundaries consistent with conformal structure are unknown.

As a first pass to studying conformal boundaries, we can examine physical data ``encoded" by the boundary. One example is the ground-state degeneracy $g$ (called the $g$-function) of the boundary state \cite{Affleck:1991tk}. This is a $c$-number that counts the ``boundary degrees of freedom." It is also associated with a thermodynamic \textit{boundary entropy} $S_{\text{b}} = \log g$ \cite{Cardy:2004hm}.

\begin{figure}[t]
\centering
\begin{tikzpicture}[yscale=0.75]
\draw[-,draw=none,fill=blue!15] (0,-1) to (0,1) to (-3,1) to (-3,-1) to (0,-1);
\node at (-1.5,0) {CFT$_{\text{L}}$};
\draw[-,draw=none,fill=red!15] (0,-1) to (0,1) to (3,1) to (3,-1) to (0,-1);
\node at (1.5,0) {CFT$_{\text{R}}$};
\draw[-,very thick,purple!60!blue] (0,-1) to (0,1);

\draw[->,thick] (2.5-0.1+1,0) to (4+0.1+1,0);
\node at (3.25+1,0.3) {Folding};

\draw[-,draw=none,fill=purple!20] (4.5+1,-1) to (4.5+1,1) to (7.5+1,1) to (7.5+1,-1) to (4.5+1,-1);
\draw[-,very thick,purple!60!blue] (4.5+1,-1) to (4.5+1,1);
\node at (6+1,0) {$\text{CFT}_{\text{L}} \otimes \overline{\text{CFT}}_{\text{R}}$};
\end{tikzpicture}
\caption{An ICFT consists of a ``left" CFT and a ``right" CFT glued together along a defect (left). This can be mapped to a BCFT (right) by folding along the defect.}
\label{figs:folding}
\end{figure}
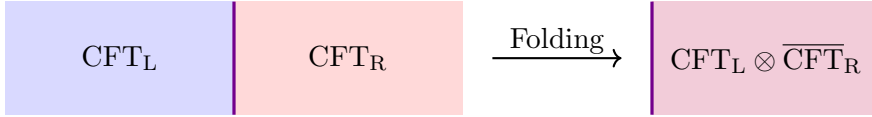

Instead of a CFT with a boundary, we may consider two CFTs glued along a defect surface (in 2d, a line) preserving (reduced) conformal symmetry. This defect is an interface between the constituent systems \cite{Wong:1994np}, so the full theory is called an \textit{interface CFT (ICFT)}. Like with boundaries, a general classification of conformal interfaces is unknown. However, an ICFT can be mapped to a boundary (B)CFT by folding along the interface (Figure \ref{figs:folding}), so these are related problems (cf. \cite{Oshikawa:1996dj}). In particular, physical parameters characterizing interfaces are encoded by boundary states, with the $g$-function being an example \cite{Azeyanagi:2007qj}.

Notably, interfaces make manifest additional physics encoded by the folded boundary states. One example is energy transport, which is characterized by \textit{transport coefficients} associated with the interface \cite{Quella:2006de,Billo:2016cpy,Meineri:2019ycm}. The proportion of energy transported across the interface is quantified by a transmission coefficient $\mathcal{T}$, while the proportion that is bounced back is described by a reflection coefficient $\mathcal{R}$. These transport coefficients are defined through expectation values of the stress tensor by \cite{Quella:2006de} and sum to $1$ by construction. Furthermore, unitarity can be used to bound them between $0$ and $1$ \cite{Billo:2016cpy,Meineri:2019ycm}.

An underlying motivation of this work is to advocate for transport coefficients as describing a facet of the physics of conformal defects apart from the $g$-function. This is a rather broad goal, so to narrow our focus we examine a particular class of CFT---\textit{symmetric orbifold theories}. These are defined by taking $N$ copies of some ``seed" theory $\mathcal{M}$ and quotienting by the permutations constituting the symmetric group $S_N$. Motivated by the recent classification of boundaries and $g$-functions in symmetric orbifolds \cite{Belin:2021nck},\footnote{See also \cite{Gaberdiel:2021kkp} for similar work in the context of string theory and AdS$_3$/CFT$_2$ holography.} our main reasons for considering these theories lie in both their tractability (given information about the seed theory) and their connections to holographic CFT (cf. \cite{Haehl:2014yla,Belin:2014fna}). The general review and analysis of symmetric orbifold CFT at the orbifold point constitutes Section \ref{sec:symmOrbi}.

Furthermore, some symmetric orbifold theories can be understood directly at strong coupling\footnote{We are referring to the marginal coupling which, when turned on, takes us away from the orbifold point \cite{Avery:2010er}. This coupling makes the $N$ copies of the theory interact.} via the AdS/CFT correspondence \cite{Maldacena:1997re}, thereby giving us access to a regime in which field-theoretic calculations are otherwise typically intractable. For example, type IIB supergravity on $\text{AdS}_3 \times S^3 \times \mathbb{T}^4$ is dual to the symmetric orbifold of a $\mathbb{T}^4$ sigma model at strong coupling and with a large number of copies, and this duality also persists in more stringy/weakly coupled regimes \cite{Eberhardt:2019ywk}. More pertinent to our purposes, we can describe a simple class of top-down holographic conformal interfaces in this CFT and at strong coupling through a non-supersymmetric\footnote{There are also supersymmetric deformations of the AdS$_3 \times S^3 \times \mathbb{T}^4$ vacuum \cite{Chiodaroli:2009yw,Chiodaroli:2010ur}.} dilatonic deformation of the AdS$_3$ bulk. The resulting gravitational background is called Janus \cite{Bak:2007jm}. By using the holographic prescription initially proposed for ``thin-brane" configurations \cite{Bachas:2020yxv} and subsequently refined by \cite{Baig:2022cnb} and \cite{Bachas:2022etu}, one may obtain transport coefficients encoded by particular boundary states in the strongly coupled sector of the ICFT dual to type IIB on Janus.

Indeed, the holographic transmission coefficients of the Janus interface have been obtained recently \cite{Bachas:2022etu}. In Section \ref{sec:t4Symm}, we will compare these coefficients against those of the Janus interface at the orbifold point, which we obtain via applying the procedure of Section \ref{sec:symmOrbi} to an appropriate boundary state of the folded $\mathbb{T}^4$ seed theory. The seed theory is four non-interacting copies of a free $S^1$-valued scalar field, each with a jump in mass along a defect. Under folding, each individual $S^1$ theory maps to two non-interacting scalar fields on half space that together are taken to satisfy a ``Neumann--Dirichlet" boundary condition \cite{Azeyanagi:2007qj}. Transport in the folded free scalar theory with this boundary condition has long been understood \cite{Quella:2006de}, but the main significance of Section \ref{sec:symmOrbi} is to emphasize the answer does not change in the symmetric orbifold. Our procedure ultimately yields an approximate answer for transport coefficients in the weakly coupled regime of the $\mathbb{T}^4$ symmetric orbifold theory.

Performing this type of strong-weak comparison is not a new application of the holographic nature of the Janus solution. \cite{Azeyanagi:2007qj} exploited the tractability of both the strongly coupled and weakly coupled regimes of the $\mathbb{T}^4$ symmetric orbifold ICFT to study how the boundary entropies $S_{\text{b}}$ of these interfaces run with coupling.\footnote{They assume the boundary entropy has no contributions from twist fields. In light of \cite{Belin:2021nck}, a boundary state agreeing with this assumption is more consistent with a bulk geometrical interpretation (i.e. a supergravity state) but is also ``atypical." We discuss this more in the Section \ref{sec:janInt}.} They found that $S_{\text{b}}$ is highly insensitive to coupling, running only a small amount. Going further, a similar calculation in supersymmetric Janus \cite{Chiodaroli:2010ur} found an exact match between the strongly coupled and weakly coupled regimes. However, in our comparison, we find that $\mathcal{T}$ changes much more nontrivially with coupling when the parameter characterizing the strength of the dilatonic deformation is not at the ends of its regime of validity (Figure \ref{figs:transComp}).

This makes sense. Heuristically, quantities describing transport are more sensitive to coupling than those describing thermodynamics. The classic example is $\mathcal{N} = 4$ supersymmetric Yang--Mills in which free energy \cite{Gubser:1998nz,Fotopoulos:1998es} only changes by a factor of $\frac{3}{4}$ while shear viscosity of the SYM plasma \cite{Policastro:2001yc,Buchel:2004di,Huot:2006ys} changes infinitely. Inspired by this story, we interpret our result as additional evidence for the idea that transport is more sensitive to coupling when compared to thermodynamics.

\section{Interface data in symmetric orbifold theories}\label{sec:symmOrbi}

We start by schematically discussing the physical data of conformal interfaces in the symmetric orbifold of a \textit{known} ICFT $\mathcal{M}$. The $N$-fold symmetric orbifold theory $\widetilde{\mathcal{M}}_N$ consists of $N$ copies of $\mathcal{M}$ modded out by the symmetric group $S_N$ of permutations of the factors:
\begin{equation}
\widetilde{\mathcal{M}}_N = \mathcal{M}^{\otimes N}/S_N.
\end{equation}
Here we focus on the ``orbifold point" of the theory, in which the $N$ copies are taken to be non-interacting. This can also be seen as the free sector of the theory.

Instead of dealing with conformal interfaces directly, it is useful to map them to conformal boundaries via folding \cite{Quella:2006de}. Suppose that the seed ICFT $\mathcal{M}$ consists of two CFTs $\mathcal{C}_{\text{L}}$ and $\mathcal{C}_{\text{R}}$ glued along an interface. We map this to a seed BCFT $\mathcal{C}_{\text{L}} \otimes \overline{\mathcal{C}}_{\text{R}}$. This induces a mapping between $\widetilde{\mathcal{M}}_N$ and $(\widetilde{\mathcal{C}_{\text{L}} \otimes \overline{\mathcal{C}}_{\text{R}}})_{N}$. Thus, we can use the technology of BCFT and boundary states to study conformal interfaces (cf. \cite{Wong:1994np,Oshikawa:1996dj}).

Following \cite{Belin:2021nck}, we start by assuming knowledge about interface (or boundary) data in the seed theory. Using this knowledge and the combinatorics of $S_N$, we then establish a recipe for interface (or boundary) data in the associated symmetric orbifold theory.

Before getting into the details, we remark that the mapping of $\widetilde{\mathcal{M}}_N$ to $(\widetilde{\mathcal{C}_{\text{L}} \otimes \overline{\mathcal{C}}_{\text{R}}})_{N}$ is subtle. Specifically, taking a symmetric orbifold of $\mathcal{M}$ then folding yields $(\widetilde{\mathcal{C}}_{\text{L}})_N \otimes \overline{(\widetilde{\mathcal{C}}_{\text{R}})_N}$, which is not technically the same as taking the symmetric orbifold of $\mathcal{C}_{\text{L}} \otimes \overline{\mathcal{C}}_{\text{R}}$. However, we insist that the latter can be used as a proxy for the former when computing data associated with transport, because the relevant CFT data is encoded by overlaps of product states.

\subsection{Twisted sectors of symmetric orbifolds}

Let us first review the basic structure of symmetric orbifold theories, following \cite{Belin:2021nck}. There are two major differences between the spectra of the $N$-fold product theory $\mathcal{M}^{\otimes N}$ and of the symmetric orbifold theory $\widetilde{\mathcal{M}}_N$. The first is that only product states that are invariant under all permutations, i.e. totally symmetric product states, survive the orbifolding procedure. These particular states constitute the so-called \textit{untwisted sector} of $\widetilde{\mathcal{M}}_N$.

The second difference is the presence of \textit{twisted sectors} in the orbifolded theory. Recall that the permutation group acts on different seed copies of $\mathcal{M}$. We can systematically construct states of the orbifold by starting with an individual permutation $\sigma \in S_N$ and identifying (``gluing together") copies of the seed theory $\mathcal{M}$ as instructed by $\sigma$. In more mathematically precise terms, such a state (defined on the cylinder) is one where the corresponding field is periodic up to the action of the permutation \cite{Burrington:2018upk,Apolo:2022fya}. Concretely, for a generic field $\varphi(z)$ ($z \in S^1$) of the seed theory, we define a field $\Phi(z)$ for which a $2\pi$ rotation ($z \to e^{2\pi i}z$) maps $\Phi(z) = \varphi_{(i)}(z)$ [where $\varphi_{(i)}$ is the $i$th copy of $\varphi$] to $\varphi_{(\sigma(i))}(z)$. Gluing yields a Hilbert space distinct from that of $\mathcal{M}^{\otimes N}$ and consisting of ``$\sigma$-twisted" states. Allowing a slight abuse of notation,\footnote{Taking a quotient technically requires a group (which $\{\sigma\}$ is not). However, we can still use the notation of quotients to describe seed theories being glued in a particular order and by a specific permutation.} we write the space as $\mathcal{M}^{\otimes N}/\{\sigma\}$.

In the full symmetric orbifold, we mod out by the full group $S_N$, so the states that survive this procedure must be invariant under all permutations. As a result, each twisted sector consists of totally symmetric sums of twisted states taken over all permutations of a particular cycle type. For example, in the $N = 3$ case, we would sum together states twisted by permutations $(1\,2), (1\,3), (2\,3) \in S_3$ (using cycle notation) to get states in one of the (symmetrized) twisted sectors, and we would similarly sum together states twisted by $(1\,2\,3),(1\,3\,2) \in S_3$ to get those of the other sector. See Figure \ref{figs:twistedSym} for a cartoon.

\begin{figure}
\centering
\subfloat[Untwisted sector]{
\begin{tikzpicture}[scale=0.7]
%\node at (-1,0.5) {\large{Sym}};
%\node[white] at (4+1,0.5) {\large{Sym}};

%\draw[-,thick] (-0.125,1.125) to (-0.25,1.125) to (-0.25,-0.125) to (-0.125,-0.125);
%\draw[-,thick] (4+0.125,1.125) to (4+0.25,1.125) to (4+0.25,-0.125) to (4+0.125,-0.125);

\draw[-,thick] (0,0) to (1,0) to (1,1) to (0,1) to (0,0);
\draw[-,thick] (1.5,0) to (2.5,0) to (2.5,1) to (1.5,1) to (1.5,0);
\draw[-,thick] (3,0) to (4,0) to (4,1) to (3,1) to (3,0);

\node at (0.5,0.5) {$\mathcal{M}^1$};
\node at (2,0.5) {$\mathcal{M}^2$};
\node at (3.5,0.5) {$\mathcal{M}^3$};

\node at (1.5,-0.5) {};
\end{tikzpicture}
}\linebreak
\subfloat[$2$-twisted sector]{
\begin{tikzpicture}[scale=0.7]
\node at (1.5,-0.5) {};

\draw[-,thick] (0-5,0) to (1-5,0) to (1-5,1) to (0-5,1) to (0-5,0);
\draw[-,thick] (1.5-5,0) to (2.5-5,0) to (2.5-5,1) to (1.5-5,1) to (1.5-5,0);
\draw[-,thick] (3-5,0) to (4-5,0) to (4-5,1) to (3-5,1) to (3-5,0);

\draw[-,thick] (0.5-5,1) arc (180:0:0.75 and 0.5);
\draw[->>,thick] (0.5-5,1) arc (180:120:0.75 and 0.5);
\draw[->>,thick] (0.5-5,1) arc (180:30:0.75 and 0.5);

\draw[-,thick] (2-5,0) arc (0:-180:0.75 and 0.5);
\draw[->>,thick] (2-5,0) arc (0:-150:0.75 and 0.5);
\draw[->>,thick] (2-5,0) arc (0:-60:0.75 and 0.5);

\node at (0.5-5,0.5) {$\mathcal{M}^1$};
\node at (2-5,0.5) {$\mathcal{M}^2$};
\node at (3.5-5,0.5) {$\mathcal{M}^3$};

\node at (-0.5,0.5) {\large$+$};

\draw[-,thick] (0,0) to (1,0) to (1,1) to (0,1) to (0,0);
\draw[-,thick] (1.5,0) to (2.5,0) to (2.5,1) to (1.5,1) to (1.5,0);
\draw[-,thick] (3,0) to (4,0) to (4,1) to (3,1) to (3,0);

\draw[-,thick] (0.5,1) arc (180:0:1.5 and 0.5);
\draw[->>,thick] (0.5,1) arc (180:135:1.5 and 0.5);
\draw[->>,thick] (0.5,1) arc (180:45:1.5 and 0.5);

\draw[-,thick] (3.5,0) arc (0:-180:1.5 and 0.5);
\draw[->>,thick] (3.5,0) arc (0:-135:1.5 and 0.5);
\draw[->>,thick] (3.5,0) arc (0:-45:1.5 and 0.5);

\node at (0.5,0.5) {$\mathcal{M}^1$};
\node at (2,0.5) {$\mathcal{M}^2$};
\node at (3.5,0.5) {$\mathcal{M}^3$};

\node at (-0.5+5,0.5) {\large$+$};

\draw[-,thick] (0+5,0) to (1+5,0) to (1+5,1) to (0+5,1) to (0+5,0);
\draw[-,thick] (1.5+5,0) to (2.5+5,0) to (2.5+5,1) to (1.5+5,1) to (1.5+5,0);
\draw[-,thick] (3+5,0) to (4+5,0) to (4+5,1) to (3+5,1) to (3+5,0);

\draw[-,thick] (0.5+5+1.5,1) arc (180:0:0.75 and 0.5);
\draw[->>,thick] (0.5+5+1.5,1) arc (180:120:0.75 and 0.5);
\draw[->>,thick] (0.5+5+1.5,1) arc (180:30:0.75 and 0.5);

\draw[-,thick] (3.5+5,0) arc (0:-180:0.75 and 0.5);
\draw[->>,thick] (3.5+5,0) arc (0:-150:0.75 and 0.5);
\draw[->>,thick] (3.5+5,0) arc (0:-60:0.75 and 0.5);

\node at (0.5+5,0.5) {$\mathcal{M}^1$};
\node at (2+5,0.5) {$\mathcal{M}^2$};
\node at (3.5+5,0.5) {$\mathcal{M}^3$};
\end{tikzpicture}
}\linebreak
\subfloat[3-twisted sector\label{figs:3twisted}]{
\begin{tikzpicture}[scale=0.7]
\node at (1.5,-0.5) {};

\draw[-,thick] (0,0) to (1,0) to (1,1) to (0,1) to (0,0);
\draw[-,thick] (1.5,0) to (2.5,0) to (2.5,1) to (1.5,1) to (1.5,0);
\draw[-,thick] (3,0) to (4,0) to (4,1) to (3,1) to (3,0);

\draw[-,thick] (0.5+0.1,1) arc (180:0:0.75-0.1 and 0.5);
\draw[->>,thick] (0.5+0.1,1) arc (180:120:0.75-0.1 and 0.5);
\draw[->>,thick] (0.5+0.1,1) arc (180:30:0.75-0.1 and 0.5);

\draw[-,thick] (0.5+1.5+0.1,1) arc (180:0:0.75-0.1 and 0.5);
\draw[->>,thick] (0.5+1.5+0.1,1) arc (180:120:0.75-0.1 and 0.5);
\draw[->>,thick] (0.5+1.5+0.1,1) arc (180:30:0.75-0.1 and 0.5);

\draw[-,thick] (3.5,0) arc (0:-180:1.5 and 0.5);
\draw[->>,thick] (3.5,0) arc (0:-135:1.5 and 0.5);
\draw[->>,thick] (3.5,0) arc (0:-45:1.5 and 0.5);

\node at (0.5,0.5) {$\mathcal{M}^1$};
\node at (2,0.5) {$\mathcal{M}^2$};
\node at (3.5,0.5) {$\mathcal{M}^3$};

\node at (-0.5+5,0.5) {\large$+$};

\draw[-,thick] (0+5,0) to (1+5,0) to (1+5,1) to (0+5,1) to (0+5,0);
\draw[-,thick] (1.5+5,0) to (2.5+5,0) to (2.5+5,1) to (1.5+5,1) to (1.5+5,0);
\draw[-,thick] (3+5,0) to (4+5,0) to (4+5,1) to (3+5,1) to (3+5,0);

\draw[-,thick] (0.5+5,1) arc (180:0:1.5 and 0.5);
\draw[->>,thick] (0.5+5,1) arc (180:135:1.5 and 0.5);
\draw[->>,thick] (0.5+5,1) arc (180:45:1.5 and 0.5);

\draw[-,thick] (3.5+5-0.1,0) arc (0:-180:0.75-0.1 and 0.5);
\draw[->>,thick] (3.5+5-0.1,0) arc (0:-150:0.75-0.1 and 0.5);
\draw[->>,thick] (3.5+5-0.1,0) arc (0:-60:0.75-0.1 and 0.5);

\draw[-,thick] (2+5-0.1,0) arc (0:-180:0.75-0.1 and 0.5);
\draw[->>,thick] (2+5-0.1,0) arc (0:-150:0.75-0.1 and 0.5);
\draw[->>,thick] (2+5-0.1,0) arc (0:-60:0.75-0.1 and 0.5);

\node at (0.5+5,0.5) {$\mathcal{M}^1$};
\node at (2+5,0.5) {$\mathcal{M}^2$};
\node at (3.5+5,0.5) {$\mathcal{M}^3$};
\end{tikzpicture}}
\caption{A cartoon of the sectors of the symmetric orbifold theory $\widetilde{\mathcal{M}}_3 = \mathcal{M}^{\otimes 3}/S_3$. Each box is a copy of the seed theory. Boxes connected by arrows are identified with one another (i.e. ``glued"), and any remaining boxes are just factors in the symmetric product. The untwisted sector (a) is simply the symmetrized product space. The 2-twisted sector (b) is constructed by gluing together any two seed copies and symmetrizing. The 3-twisted sector (c) arises from gluing together three copies in any order and symmetrizing.}
\label{figs:twistedSym}
\end{figure}

Note that the permutations of a particular cycle type precisely constitute a particular conjugacy class of $S_N$. Through this equivalence, the conjugacy classes each correspond to an integer partition of $N$ (the cycle type). So, the number of twisted sectors in the symmetric orbifold theory matches the number of distinct integer partitions of $N$.

\paragraph{Twisted states}

Let us be more explicit about the properties of twisted states. First for simplicity, consider a single $N$-cycle $\sigma_N \in S_N$, by which we mean that $\sigma_N$ is an order $N$ ($|\sigma_N| = N$) connected permutation. In other words, $\sigma_N$ has no fixed points or subcycles. We take the $N$ copies of the seed theory permuted by $\sigma_N$ and glue them in the order dictated by $\sigma_N$. For example, taking $N = 3$, we have two 3-cycles, $(1\,2\,3)$ and $(1\,3\,2)$, and the corresponding gluings of seed theories are respectively visualized in Figure \ref{figs:3twisted}.

For some seed primary scalar $\ket{h}$ of weight $h$, the Hilbert space of $\mathcal{M}^{\otimes N}/\{\sigma_N\}$ has a state $\ket{\sigma_N \cdot h^{(N)}}$ of weight $\mathfrak{h}\big[h^{(N)}\big]$ (using $\mathfrak{h}$ to represents weights in the orbifold), where
\begin{equation}
\mathfrak{h}\big[h^{(N)}\big] \equiv \frac{c}{24}\left(N - \frac{1}{N}\right) + \frac{h}{N}.\label{twistedPrimaryk}
\end{equation}
The superscript $(N)$ is meant to highlight that we are starting with $N$ copies of the seed state $\ket{h}$ priori to gluing them by $\sigma_N$. Note that the explicit choice of $\sigma_N$ does not matter.

This is still not a state of the symmetric orbifold, since it is not invariant under all permutations. Instead, we define the symmetrization of $\ket{\sigma_N \cdot h^{(N)}}$ as
\begin{equation}
\ket{h^{(N)}} \equiv \frac{1}{\sqrt{(N-1)!}} \sum_{\sigma \in \{N\text{-cycles}\}} \ket{\sigma \cdot h^{(N)}}.\label{maxtwisted}
\end{equation}
Basically, we are summing over all possible $N$-cycles. This is mathematically equivalent to summing over all elements of $S_N$ in the same conjugacy class as a particular $\sigma_N$. There are $(N-1)!$ such permutations, and so $\frac{1}{\sqrt{(N-1)!}}$ is a normalization factor for the symmetrized state. Furthermore, this can be rewritten as a sum over all possible conjugates of $\sigma_N$ in the symmetric group, so long as we introduce a $\frac{1}{N}$ factor to prevent overcounting (cf. \cite{Lunin:2000yv,Burrington:2018upk}):
\begin{equation}
\ket{h^{(N)}} = \frac{1}{N\sqrt{(N-1)!}} \sum_{\tau \in S_N} \ket{(\tau \sigma_N \tau^{-1}) \cdot h^{(N)}}.\label{twistedfullfin}
\end{equation}
This rewriting can be found using finite group theory. We first explicitly write each distinct $N$-cycle in the sum as some conjugate of $\sigma_N$. We then further rewrite each term as a sum over the centralizer of $\sigma_N$ [defined as $\mathcal{S}(\sigma_N) \equiv \{g \in S_N\,|\,g\sigma_N g^{-1} = \sigma_N\}$] by the replacement $\sigma_N \to \frac{1}{N}\sum_{g \in \mathcal{S}(\sigma_N)} g\sigma_N g^{-1}$ (where the factor of $N$ is the size of the centralizer and can be found through the orbit-stabilizer theorem). We then use the fact that the left cosets of $\mathcal{S}(\sigma_N)$ in $S_N$ partition the symmetric group to get the right-hand side of \eqref{twistedfullfin}.

The fact that $\ket{h^{(N)}}$ can be written as a sum over the full symmetric group means that it is manifestly invariant under permutations. So, it is also in the Hilbert space of the symmetric orbifold theory $\widetilde{\mathcal{M}}_N$ despite not being a product state, and it still has the same weight as its individual terms as given by \eqref{twistedPrimaryk}. We will refer to the states of the form $\ket{h^{(N)}}$ as ``maximally twisted" in that they are constructed by gluing together all $N$ copies of the seed theory.

We can generalize the above construction to arbitrary elements of $S_N$ to go beyond just $N$-cycles. Specifically, consider a permutation $\sigma = \sigma_1 \cdots \sigma_m \in S_N$, where each $\sigma_i$ is a $k_i$-cycle (with $k_i \leq k_{i+1}$ for all $i$) and the different factors are disjoint. Every element of $S_N$ can be written uniquely in this way for some $k_i \geq 1$ for which $\sum_{i=1}^m k_i = N$, and the ordered multiset $\{k_1,\dots, k_m\}$ is the \textit{cycle type} of $\sigma$. From $m$ seed primaries $\ket{h_1},\dots,\ket{h_m}$, we construct a $\sigma$-twisted state denoted by
\begin{equation}
\ket{\sigma_1 \cdot h_1^{(k_i)}} \cdots \ket{\sigma_m \cdot h_m^{(k_m)}} \in \mathcal{M}^{\otimes n}/\{\sigma\}.\label{productsigmatwist}
\end{equation}
This notation unambiguously encodes both the numbering of the seed states and how identical copies of such states are twisted together, since the constituent disjoint cycles encode the initial ``locations" of the seed states within the parent tensor state.\footnote{For example, $\ket{(1)\cdot h_1^{(1)}} \ket{(2\,3) \cdot h_2^{(2)}}$ describes a state for which we take the tensor product $\ket{h_1} \otimes \ket{h_2} \otimes \ket{h_2}$ and then glue the second and third seed factors by quotienting by $(1)(2\,3)$.} Additionally, the weight of the generic twisted state \eqref{productsigmatwist} is
\begin{equation}
\mathfrak{h}\big[h_1^{(k_i)},\dots,h_m^{(k_m)}\big] \equiv \sum_{i=1}^m \mathfrak{h}\big[h_i^{(k_i)}\big] = \sum_{i=1}^m \left[\frac{c}{24}\left(k_i - \frac{1}{k_i}\right) + \frac{h_i}{k_i}\right],
 \label{sumprimariesTwists}
\end{equation}
which we get by noting that \eqref{productsigmatwist} is basically an $m$-fold product of maximally twisted states with respect to $\mathcal{M}^{\otimes k_1},\dots,\mathcal{M}^{\otimes k_m}$. Indeed, if we consider a state twisted by an $N$-cycle ($m = 1$ and $k_1 = N$) then we recover \eqref{twistedPrimaryk}, whereas taking the identity element ($m = N$ and $k_1 = \cdots = k_N = 1$) yields the weight ($h_1 + \cdots + h_N$) of a product state in $\mathcal{M}^{\otimes N}$.

Just as in the maximally twisted case above, however, we need to symmetrize the $\sigma$-twisted state in order to get a permutation-invariant state of the full symmetric orbifold theory. Again, we define a sum over all permutations $\sigma \in S_N$ of a particular cycle type:
\begin{align}
\ket{h_1^{(k_1)}} \odot \cdots \odot \ket{h_m^{(k_m)}} &\equiv \frac{1}{\sqrt{\mathcal{C}(k_1,\dots,k_m)}} \sum_{\substack{\sigma = \sigma_1 \cdots \sigma_m\\|\sigma_i| = k_i}} \ket{\sigma_1 \cdot h_1^{(k_1)}} \cdots \ket{\sigma_m \cdot h_m^{(k_m)}},\nonumber\\
&\qquad\text{where}\ \left.\sigma_i \cap \sigma_j\right|_{i \neq j} = \varnothing,\ \ k_1 + \cdots + k_m = N,
\end{align}
where we use $\odot$ to emphasize that the factors $\ket{h_i^{(k_i)}}$ of the symmetrized state on the left-hand side \textit{commute} by construction. As in \eqref{maxtwisted}, we introduce a normalization factor $\mathcal{C}(k_1,\dots,k_m)$ that is equivalent to the number of terms in the sum. Note that $\mathcal{C}(k_1,\dots,k_m)$ is just the size of the conjugacy class if $k_i = k_j \implies i = j$ (as in the maximally twisted case), but if there are multiple subcycles of the same size then this is no longer true (as in the untwisted case for which $\mathcal{C} = N!$). For simplicity, we will ignore this normalization unless working with specific cases.

Most importantly, through the same group-theoretic techniques described in the maximally twisted case, we can fix a particular $\{k_1,\dots,k_m\}$-cycle $\sigma$ and rewrite the sum as one over the full symmetric group (omitting the cycle-type-dependent normalization):
\begin{align}
\ket{h_1^{(k_1)}} \odot \cdots \odot \ket{h_m^{(k_m)}} &\propto \sum_{\tau \in S_N} \ket{\sigma_1' \cdot h_1^{(k_1)}} \cdots \ket{\sigma_m' \cdot h_m^{(k_m)}},\nonumber\\
&\qquad\text{where}\ \tau \sigma \tau^{-1} = \sigma_1' \cdots \sigma_m',\ \ |\sigma_i'| = k_i,
\label{twistedgeneric}
\end{align}
The key takeaway of \eqref{twistedgeneric} is that $\ket{h_1^{(k_1)}} \odot \cdots \odot \ket{h_m^{(k_m)}}$ is a sum over $S_N$ and thus invariant under permutations. This construction also explicitly shows how different sectors of $\mathcal{M}^{\otimes N}/S_N$ are labeled by the cycle types, which in turn label conjugacy classes of $S_N$.

As a concrete example of these twisted states, consider again $N = 3$, and take two seed primaries $\ket{h_1}$ and $\ket{h_2}$ with one copy of the former and two copies of the latter. We can use these primaries to build states respectively twisted by $(1\,2)$, $(1\,3)$, or $(2\,3)$:
\begin{align}
\ket{(1)\cdot h_1} \ket{(2\,3)\cdot h_2} &\in \mathcal{M}^{\otimes 3}/\{(1)(2\,3)\},\\
\ket{(2)\cdot h_1} \ket{(1\,3)\cdot h_2} &\in \mathcal{M}^{\otimes 3}/\{(2)(1\,3)\},\\
\ket{(3)\cdot h_1} \ket{(1\,2)\cdot h_2} &\in \mathcal{M}^{\otimes 3}/\{(3)(1\,2)\}.
\end{align}
All three of these states have the same weight,
\begin{equation}
\left[\frac{c}{24}\left(1 - \frac{1}{1}\right) + \frac{h_1}{1}\right] + \left[\frac{c}{24} \left(2 - \frac{1}{2}\right) + \frac{h_2}{2}\right] = \frac{c}{16} + h_1 + \frac{h_2}{2}.
\end{equation}
The permutation-invariant state with this weight and consisting of $\mathcal{C}(1,2) = 3$ terms is
\begin{equation}
\begin{split}
\ket{h_1^{(1)}} \odot \ket{h_2^{(2)}} = \frac{1}{\sqrt{3}} \big[\ket{(1)\cdot h_1} \ket{(2\,3)\cdot h_2} &+ \ket{(2)\cdot h_1} \ket{(1\,3)\cdot h_2}\\
& +  \ket{(3)\cdot h_1} \ket{(1\,2)\cdot h_2}\big].
\end{split}
\end{equation}
This is indeed invariant under permutations and thus is a well-defined state of the symmetric orbifold, despite not being a product state.

We should mention that each twisted sector has its own ground state realized as a twist of the seed vacuum state $h = 0$. These states are called \textit{bare twists}. In terms the notation of \eqref{twistedgeneric} above, they are written as $\ket{0^{(k_1)}} \odot \cdots \odot \ket{0^{(k_m)}}$. For instance, consider again a single $N$-cycle $\sigma_N$. From \eqref{twistedPrimaryk}, the weight of the twist of the vacuum by $\sigma_N$ is
\begin{equation}
\mathfrak{h}\big[0^{(N)}\big] = \frac{c}{24} \left(N - \frac{1}{N}\right),
\end{equation}
which is below the black-hole threshold $\frac{c}{24}$ and is also the weight of the (symmetrized) bare twist $\ket{0^{(N)}}$. Weights of bare twists by more complicated cycle types can similarly be computed with \eqref{sumprimariesTwists}, but they are all upper-bounded by $\mathfrak{h}\big[0^{(N)}\big]$.

As an example, consider the $N = 4$ case. There are five different cycle types in $S_4$ corresponding to the integer partitions of $4$. The associated bare-twist weights are
\begin{equation}
\begin{matrix*}[l]
\mathfrak{h}\big[0^{(1)},0^{(1)},0^{(1)},0^{(1)}\big] = 0,
&&\mathfrak{h}\big[0^{(1)},0^{(1)},0^{(2)}\big] = \dfrac{c}{16},\vspace{0.15cm}\\
\mathfrak{h}\big[0^{(1)},0^{(3)}\big] = \dfrac{c}{9},
&&\mathfrak{h}\big[0^{(2)},0^{(2)}\big] = \dfrac{c}{8},
&&\mathfrak{h}\big[0^{(4)}\big] = \dfrac{5c}{32}.
\end{matrix*}
\end{equation}
Note the untwisted sector furnishes the true vacuum of the theory $\ket{0}^{\otimes N}$ of weight $\mathfrak{h} = 0$.

\paragraph{Symmetry structure} Take the holomorphic and antiholomorphic Virasoro symmetry of the seed theory, collectively denoted as $\text{Vir}(\mathcal{M}) \oplus \overline{\text{Vir}}(\mathcal{M})$ and with central charge $c$. The algebra of the $i$th copy is generated by $\big\{L^{(i)}_n \in \text{Vir}(\mathcal{M})\big\}$ and $\big\{\overline{L}^{(i)}_n \in \overline{\text{Vir}}(\mathcal{M})\big\}$, where
\begin{equation}
\begin{split}
[L^{(i)}_n,L_{n'}^{(i)}] &= (n-n')L^{(i)}_{n+n'} + \frac{c}{12}n(n^2-1)\delta_{n+n',0},\\
[\overline{L}^{(i)}_n,\overline{L}^{(i)}_{n'}] &= (n-n')\overline{L}^{(i)}_{n+n'} + \frac{c}{12}n(n^2-1)\delta_{n+n',0},\\
[L^{(i)}_n,\overline{L}^{(i)}_{n'}] &= 0.
\end{split}
\end{equation}
Now consider just the chiral symmetry algebra $\mathcal{W}$ of the seed theory. $\mathcal{W}$ generically contains $\text{Vir}(\mathcal{M})$. Meanwhile, the symmetric orbifold theory $\widetilde{\mathcal{M}}_N$ inherits the chiral algebra
\begin{equation}
\mathcal{W}^{\otimes N}/S_N \supset \text{Vir}(\mathcal{M})^{\otimes N}/S_N.\label{fullAlgSymm}
\end{equation}
Notably, this contains the chiral Virasoro symmetry $\text{Vir}(\mathcal{M}^{\otimes N})$ of the product theory. We refer to this subalgebra as the ``full" Virasoro algebra of $\widetilde{\mathcal{M}}_N$ and denote its generators by $\big\{L_n \equiv \sum_{i=1}^N L_n^{(i)}\big\}$ [resp. $\big\{\overline{L}_n \equiv \sum_{i=1}^N \overline{L}_n^{(i)}\big\}$ for the generators of the analogous antiholomorphic copy $\overline{\text{Vir}}(\mathcal{M}^{\otimes N})$]. The associated stress tensor of $\mathcal{M}^{\otimes N}$ whose modes generate the full Virasoro algebra is found by summing together the stress tensors of each copy of the seed theory. Since the product theory is non-interacting (i.e. seed Virasoro generators acting on different factors commute), the commutator of the full Virasoro generators is
\begin{equation}
[L_n,L_{n'}] = \sum_{i=1}^N [L_n^{(i)},L_{n'}^{(i)}] = (n-n')L_{n+n'} + \frac{Nc}{12}n(n^2-1)\delta_{n+n',0},
\end{equation}
with an equivalent expression for the antiholomorphic generators. Thus, the central charge of the full Virasoro algebra is $N c$.

However, note that the chiral algebra $\text{Vir}(\mathcal{M})^{\otimes N}/S_N$ consists of more than just full Virasoro generators. Notably, there are also ``fractional" Virasoro generators that act on twisted sectors of the theory \cite{Burrington:2018upk,Burrington:2022dii,Burrington:2022rtr}. To be concrete, consider the twisted sector corresponding to the cycle type $\{k_1,\dots,k_m\}$. Restricting our attention to one of the cycles of length $k_i$, the associated glued copies each furnish one copy of the seed stress tensor $T^{(j)}$ ($j = 1,\dots,k_i)$. By looking at the covering space \cite{Burrington:2018upk}, we can define fractional modes of the stress tensor $\{\ell_{n/{k_i}}\}$ by the equation
\begin{equation}
\ell_{n/{k_i}} \equiv \frac{1}{2\pi i} \oint  dz\,z^{n/k_i + 1} \sum_{j=1}^{k_i} e^{-2\pi i n(j-1)/k_i} T^{(j)}(z).\label{fracgen}
\end{equation}
These and their antiholomorphic counterparts $\{\overline{\ell}_{n/{k_i}}\}$ together satisfy the following commutation relations:
\begin{equation}
\begin{split}
[\ell_{n/k_i},\ell_{n'/k_i}] &= \left(\frac{n-n'}{k}\right)\ell_{(n+n')/k_i} + \frac{k_i c}{12} \left(\frac{n}{k_i}\right)\left[\left(\frac{n}{k_i}\right)^2 - 1\right]\delta_{n+n',0},\\
[\overline{\ell}_{n/k_i},\overline{\ell}_{n'/k_i}] &= \left(\frac{n-n'}{k_i}\right)\overline{\ell}_{(n+n')/k_i} + \frac{k_ic}{12} \left(\frac{n}{k_i}\right)\left[\left(\frac{n}{k_i}\right)^2 - 1\right]\delta_{n+n',0},\\
[\ell_{n/k_i},\overline{\ell}_{n'/k_i}] &= 0.
\end{split}\label{fracVirModes}
\end{equation}
This looks like the usual Virasoro algebra, but with fractional modes. Indeed, \eqref{fracgen} and \eqref{fracVirModes} respectively reduce to the usual integer-moded Virasoro operators and algebra $\text{Vir}(\mathcal{M}^{\otimes k_i}) \oplus \overline{\text{Vir}}(\mathcal{M}^{\otimes k_i})$ with central charge $k_i c$ when taking modes $\frac{n}{k_i} \in \mathbb{Z}$. Additionally, the fractional and full Virasoro generators share a nontrivial commutation bracket \cite{Burrington:2022dii}:
\begin{equation}
[L_n,\ell_{n'/k_i}] = [\ell_{nk_i/k_i},\ell_{n'/k_i}].
\end{equation}
The generators built from the fractional modes that act on a generic state are of the form
\begin{equation}
L_{n_1/k_1,\dots,n_m/k_m} \equiv \sum_{i=1}^m \ell_{n_i/k_i}.\label{gengenerators}
\end{equation}
By taking generators of the untwisted sector ($m = N$, $k_1 = \cdots = k_N = 1$) and summing over identical modes ($n_1 = \cdots = n_N \equiv n$), we recover the full Virasoro generators $L_n$.

The fractional Virasoro generators act on states in nontrivial twisted sectors. Hence, twisted-sector primaries and their fractional descendants may generically appear as terms in the boundary states of a symmetric orbifold theory \cite{Belin:2021nck}, as we review in Section \ref{sec:bdrystates}.

However, the boundary data with which we are primarily concerned do not involve these modes directly. Specifically, $g$-functions \cite{Affleck:1991tk} are identified as overlaps between boundary states and the untwisted vacuum \cite{Cardy:2004hm}, while the transport coefficients of the stress tensor \cite{Quella:2006de} are equated to overlaps of boundary states with full Virasoro descendents of the untwisted vacuum. Twisted-sector terms are projected out when defining either quantity. While we can extract twisted-sector data encoded by a boundary state, such as overlaps with bare twists or fractional Virasoro descendents, we will only briefly discuss such quantities in Section \ref{sec:twistedData}.

\subsection{Building the boundary states}\label{sec:bdrystates}

Via folding, all physical data about a conformal interface is encoded by an associated boundary state in the closed-string spectrum of the folded BCFT \cite{Cardy:1984bb,Cardy:2004hm}. So, a good starting point for studying transport coefficients in a symmetric orbifold theory is to explore how its boundary states might be written in terms of those of the seed theory.

Let us first clarify some fundamental statements of BCFT. A generic CFT boundary state $\ket{B}$ must satisfy the gluing condition
\begin{equation}
\left(L_n - \overline{L}_{-n}\right)\ket{B} = 0,\ \ \forall n \in \mathbb{Z},\label{ishibashiVirasoro}
\end{equation}
where we recall that $\{L_n\}$ and $\{\overline{L}_{n}\}$ are respectively the holomorphic and antiholomorphic generators of the full Virasoro symmetry. In addition, boundary states must satisfy the so-called Cardy condition \cite{Cardy:1984bb,Cardy:2004hm} obtained by equating the open-string and closed-string slicings of the 2d cylinder partition function. To classify the boundary states, one may start by finding a basis for the solution space of \eqref{ishibashiVirasoro}; the elements of such a basis are called \textit{Ishibashi states} \cite{Ishibashi:1988kg}. We may then use the Cardy conditions to compute the specific linear combinations of Ishibashi states that correspond to boundary states.

However, the full space of Ishibashi states satisfying \eqref{ishibashiVirasoro} is not generically known when the theory is not Virasoro-diagonal, and this obstructs simple attempts to organize boundary states. A more attainable goal is to classify boundary states that also respect the chiral extension that diagonalizes the CFT \cite{Ishibashi:1988kg,Blumenhagen2009}. Mathematically, this means that we restrict our attention to a smaller space of boundary states that satisfy
\begin{equation}
\big[W_n - (-1)^{s(W)}\Omega(\overline{W}_{-n})\big]\ket{B} = 0,\ \ \forall n \in \mathbb{Z},\label{genishibashi}
\end{equation}
where $\{W_n\}$ and $\{\overline{W}_n\}$ are respectively holomorphic and antiholomorphic generators of the extended symmetry algebra corresponding to a holomorphic current $W$ of spin $s(W)$, and $\Omega$ is an automorphism of the extended algebra. \eqref{genishibashi} is a generalization of the condition \eqref{ishibashiVirasoro}, since the holomorphic current $T$ corresponding to the Virasoro modes has $s(T) = 2$. The solution space of the set of constraints \eqref{genishibashi} over all extended symmetry generators can be described by Ishibashi states \cite{Ishibashi:1988kg}.

The goal of \cite{Belin:2021nck} (see also related earlier work \cite{Recknagel:2002qq}) is to construct boundary states that respect the chiral algebra $\text{Vir}(\mathcal{M})^{\otimes N}/S_N$ universal to symmetric orbifold theories. The main idea is to classify the solution space of \eqref{genishibashi}, where the generators are of the form \eqref{gengenerators}. Basically, \cite{Belin:2021nck} constructs ``twisted" Ishibashi states from twisted primaries and fractional descendants that, in conjunction with the expected ``untwisted" Ishibashi states (i.e. symmetric products of seed Ishibashi states), act as building blocks for the boundary states. For our purposes of understanding the physical data encoded by boundary states, it is instructive to review this construction. We do so now.

As a caveat, we stress that we focus on Virasoro-diagonal seed theories for concreteness, thereby assuming that the only symmetry generators that we need to worry about are those built from the seed stress tensor. However, note that the basic construction works more generally at the cost of completeness in the classification of boundary states.

\paragraph{Untwisted building blocks}

We first suppose the seed theory $\mathcal{M}$ has boundary states $|b_\alpha\rangle$ where $\alpha = 1,\dots,n_b$ is an index. $n_b$ is either finite (e.g. for a rational seed theory) or formally infinite (e.g. for an irrational seed theory). These states live in the solution space of the following constraint [denoting $i$th seed Virasoro modes by $L_n^{(i)}$ and $\overline{L}_n^{(i)}$]:
\begin{equation}
\big(L^{(i)}_n - \overline{L}^{(i)}_{-n}\big)|b_\alpha\rangle = 0,\ \ \forall n \in \mathbb{Z}.\label{bcseedcond}
\end{equation}
If the seed theory is diagonal with respect to the Virasoro algebra, then we can explicitly construct the seed boundary states because the solution space of \eqref{bcseedcond} is spanned by the Ishibashi states of $\mathcal{M}$ \cite{Ishibashi:1988kg,Onogi:1988qk}. Each of these Ishibashi states is constructed from some spinless primary of weight $h$ as follows:
\begin{align}
\kket{h} = \sum_{\mathfrak{m}} |h,\mathfrak{m}\rangle \otimes \overline{|h,\mathfrak{m}\rangle},\label{seedIshibashi}
\end{align}
where $\mathfrak{m}$ is either the empty set or an ordered multiset of positive integers ($\{m_1,\dots,m_\kappa\}$ with $i < j \implies m_i \leq m_j$). The state $|h,\mathfrak{m}\rangle$ (resp. $\overline{|h,\mathfrak{m}\rangle}$) is a holomorphic (resp. antiholomorphic) descendent of the associated primary, with the multiset labeling both the number and type of creation operators employed:\footnote{$\mathfrak{m} = \varnothing$ corresponds to the primary state itself.}
\begin{equation}
\begin{split}
|h,\{m_1,\dots,m_\kappa\}\rangle &= \left(\prod_{i=1}^\kappa L_{-m_i}^{(i)}\right)|h\rangle,\\
\overline{|h,\{m_1,\dots,m_\kappa\}\rangle} &= \left(\prod_{i=1}^\kappa\overline{L}_{-m_i}^{(i)}\right)|h\rangle.
\end{split}
\end{equation}
As \eqref{seedIshibashi} defines Ishibashi states of the seed theory, all seed boundary states are specific linear combinations of these states:
\begin{equation}
\ket{b_\alpha} = \sum_{h} \beta_{h,\alpha} \kket{h}.\label{bstateIshibashiseed}
\end{equation}
Note that the coefficients $\beta_{h,\alpha}$ take on specific values such that $\ket{b_\alpha}$ satisfies both \eqref{bcseedcond} and the Cardy condition \cite{Cardy:1984bb,Cardy:2004hm}.

In the untwisted sector of the symmetric orbifold theory, a generic symmetry operator consists of sums of seed Virasoro modes that individually act on different copies. The generic gluing condition \eqref{genishibashi} with such symmetry operators is solved by symmetric products of seed Ishibashi states, a fact that can be seen explicitly from the seed gluing condition \eqref{bcseedcond}. We refer to such symmetric products as untwisted Ishibashi states, and we observe that they are building blocks for states of the form (neglecting the normalization)
\begin{equation}
\ket{B}_{\text{untw}} \equiv \ket{b_{\alpha_1}} \odot \cdots \odot \ket{b_{\alpha_N}} \propto \sum_{\sigma \in S_N} |b_{\sigma(\alpha_1)}\rangle \otimes \cdots \otimes |b_{\sigma(\alpha_N)}\rangle,\label{untwState}
\end{equation}
which are boundary states of the symmetric product theory.

If all $N$ of the seed factors $|b_{\alpha_i}\rangle$ are distinct, then \eqref{untwState} is also the only way they can be combined in the symmetric orbifold. However, we may have degenerate factors, in which case we can also consider twists of these seed boundary states. This is an expected characteristic of orbifold theories \cite{Billo:2000yb}.

\paragraph{Twisted building blocks} We now discuss the twisted Ishibashi states of \cite{Belin:2021nck}. First, consider the maximally twisted sector, in which we have fractional Virasoro modes $\ell_{n/N}$ and $\overline{\ell}_{n/N}$ for all $n \in \mathbb{Z}$. Using these in the general gluing condition \eqref{genishibashi}, we get the constraints
\begin{equation}
\big(\ell_{n/N} - \overline{\ell}_{-n/N}\big)\ket{B} = 0,\ \ \forall n \in \mathbb{Z}.\label{maxTwConstraint}
\end{equation}
We can write Ishibashi states that solve this equation in a manner similar to the usual construction. From a maximally twisted primary $\ket{h^{(N)}}$, the associated Ishibashi state is
\begin{equation}
\kket{h^{(N)}} \equiv \sum_{\mathfrak{m}} |h^{(N)},\mathfrak{m}\rangle \otimes\overline{|h^{(N)},\mathfrak{m}\rangle},\label{twistIshi}
\end{equation}
where this time we define
\begin{equation}
\begin{split}
|h^{(N)},\{m_1,\dots,m_\kappa\}\rangle &= \left(\prod_{i=1}^\kappa \ell_{-m_i/N}\right)\ket{h^{(N)}},\\
\overline{|h^{(N)},\{m_1,\dots,m_\kappa\}\rangle} &= \left(\prod_{i=1}^\kappa\overline{\ell}_{-m_i/N}\right)\ket{h^{(N)}}.
\end{split}\label{fracdefishi}
\end{equation}
Now, for some seed boundary state $\ket{b_{\alpha}}$ whose expansion in terms of seed Ishibashi states is given by \eqref{bstateIshibashiseed}, we can write an associated maximally twisted state
\begin{equation}
\ket{b_\alpha^{(N)}} \equiv \sum_h \beta_{h,\alpha}\kket{h^{(N)}}.
\end{equation}
This contributes to boundary states of the symmetric orbifold built from $N$ identical copies of $\ket{b_\alpha}$, and so the maximally twisted Ishibashi states are also valid building blocks.

It is straightforward to generalize the above construction to arbitrary twisted sectors corresponding to cycle type $\{k_1,\dots,k_m\}$. We can construct Ishibashi states defined with respect to states twisted along $k_i$ copies (denoted as $\kket{h^{(k_i)}}$) by using the appropriate fractional Virasoro modes $\ell_{n/k_i}$ and $\overline{\ell}_{n/k_i}$ in \eqref{fracdefishi}. We then write symmetrized products,
\begin{equation}
\kket{h_1^{(k_1)}} \odot \cdots \odot \kket{h_m^{(k_m)}}.
\end{equation}
Given a list of $m$ seed boundary states $\{\ket{b_{\alpha_1}},\dots,\ket{b_{\alpha_m}}\}$, these twisted Ishibashi states are a basis for states of the form
\begin{equation}
\ket{B}_{\text{twst}} \equiv \ket{b_{\alpha_1}^{(k_1)}} \odot \cdots \odot \ket{b_{\alpha_m}^{(k_m)}}.\label{twState}
\end{equation}
\cite{Belin:2021nck} discusses further how \eqref{untwState} and \eqref{twState} can be used as building blocks for a broad class of boundary states in the symmetric orbifold theory labeled by representations of permutation subgroups in $S_N$, where these subgroups are those which permute identical copies of seed boundary states. The coefficients of associated Ishibashi states can be found through the Cardy condition. However, for our purposes it is sufficient to know only the general form of the boundary states---as linear combinations of \eqref{untwState} and \eqref{twState} weighted by characters of symmetric-group representations.

We reiterate that the basic construction does not actually need a diagonal seed theory. This is only a useful assumption to make because it affords us the Ishibashi states \eqref{seedIshibashi} as a complete basis for seed boundary states \cite{Ishibashi:1988kg}. For a non-diagonal seed theory, the idea of \cite{Belin:2021nck} to start with seed boundary states directly and sum over permutations and twists is still valid in the construction of boundary states that respect the chiral algebra $\text{Vir}(\mathcal{M})^{\otimes N}/S_N$. However, this may no longer provide a complete classification.

We also emphasize that there are other boundary states not included within this classification. In particular, we can consider just the boundary states that respect the full Virasoro algebra inherited from the symmetric product theory $\text{Vir}(\mathcal{M}^{\otimes N}) \subset \text{Vir}(\mathcal{M})^{\otimes N}/S_N$, i.e. those that only satisfy the gluing condition \eqref{ishibashiVirasoro} but not necessarily \eqref{genishibashi} for all of the symmetry generators. This would be a much more unconstrained problem that would allow for many more boundary states (cf. the case of the free boson \cite{Gaberdiel:2002my}).

\subsection{Transport coefficients from seed data}\label{sec:transportSeed}

Now, we discuss how the transport coefficients of \cite{Quella:2006de} are encoded by the boundary states constructed from the building blocks \eqref{untwState} and \eqref{twState}. The $g$-function has already been addressed by \cite{Belin:2021nck}, and we can look to that story for inspiration. Again, for simplicity we assume that the seed theory has no extended symmetry beyond Virasoro.

The $g$-function of a boundary state $\ket{B}$ is defined as the overlap between the boundary state and the vacuum $\ket{0}^{\otimes N}$ \cite{Affleck:1991tk}. Mathematically,
\begin{equation}
g(\ket{B}) \equiv \bra{0}^{\otimes N} \ket{B}.\label{gfunc}
\end{equation}
However, the state $\ket{0}^{\otimes N}$ lives in the untwisted sector, and so it projects out any twisted-sector terms in $\ket{B}$. As a result, for $\ket{B}$ built from seed boundary states $\ket{b_{\alpha_1}},\dots,\ket{b_{\alpha_N}}$, the $g$-function is merely the vacuum overlap with \eqref{untwState} up to an overall normalization factor, so it is proportional to a product of seed $g$-functions:
\begin{equation}
g(\ket{B}) = \mathcal{F}(N,\ket{B}) \left(g_{\alpha_1} \cdots g_{\alpha_N}\right),\ \ g_{\alpha} \equiv \bra{0}{b_{\alpha}}\rangle,\label{gfuncSymOrbi}
\end{equation}
The normalization factor of $\ket{B}$ depends on both $N$ and the number of twisted-sector seed terms that can be and are included. This carries over to the coefficient $\mathcal{F}$, and so the possibility of twisted-sector terms can influence $g$ indirectly.

Now, we discuss transport. For any 2d ICFT prior to folding, we have a ``left" theory whose full Virasoro generators we write as $\{L_n^{\text{[L]}},\overline{L}_n^{\text{[L]}}\}$ and a ``right" theory with generators $\{L_n^{\text{[R]}},\overline{L}_n^{\text{[R]}}\}$ (note the square brackets). These theories also respectively have central charges $c_{\text{L}}$ and $c_{\text{R}}$, respectively.\footnote{We assume that the holomorphic and antiholomorphic data of the two theories are the same, e.g. $c_{\text{L}} = \bar{c}_{\text{L}}$ and $c_{\text{R}} = \bar{c}_{\text{R}}$.} Upon folding, \cite{Quella:2006de} defines a $2 \times 2$ ``transport matrix"
\begin{equation}
R_{IJ}(\ket{B}) = \frac{\bra{\Omega}L_2^{[I]} \overline{L}_2^{[J]}\ket{B}}{\bra{\Omega}{B}\rangle},\ \ I,J = \text{L},\text{R},\label{matDefGen}
\end{equation}
where $\ket{\Omega} = \ket{\Omega_{\text{L}}} \otimes \ket{\Omega_{\text{R}}}$ is the vacuum of the generic folded theory. In terms of the transport matrix, \cite{Quella:2006de} defines the associated ``transmission" and ``reflection" coefficients as
\begin{equation}
\mathcal{T}(\ket{B}) = \frac{2}{c_{\text{L}}+c_{\text{R}}}\left[R_{\text{L}\text{R}}(\ket{B}) + R_{\text{R}\text{L}}(\ket{B})\right],\ \ \mathcal{R} = \frac{2}{c_{\text{L}}+c_{\text{R}}}\left[R_{\text{L}\text{L}}(\ket{B}) + R_{\text{R}\text{R}}(\ket{B})\right].\label{transRefGen}
\end{equation}
We can use the gluing conditions \eqref{ishibashiVirasoro} and the Virasoro algebra to show that $R_{IJ}$ is symmetric. Thus $\mathcal{T}$ as defined above only has one parameter (not including the central charges $c_{\text{L}},c_{\text{R}}$), which we write using the notation of \cite{Meineri:2019ycm}:
\begin{equation}
c_{\text{LR}}(\ket{B}) \equiv R_{\text{LR}}(\ket{B}) + R_{\text{RL}}(\ket{B}) = 2R_{\text{LR}}(\ket{B}).\label{clrdefBdry}
\end{equation}
In fact, $\mathcal{T} + \mathcal{R} = 1$, and so $c_{\text{LR}}$ is the only free parameter needed to specify the coefficients \eqref{transRefGen}. Furthermore, the analogous matrices defined with higher-level Virasoro modes do not give additional information because by the gluing condition \eqref{ishibashiVirasoro} (taking $n \geq 2$)
\begin{equation}
\begin{split}
&\bra{\Omega}L_n^{[I]} \overline{L}_{n+1}^{[J]} \big[L_1^{\text{[L]}} + L_1^{\text{[R]}} - \overline{L}_{-1}^{\text{[L]}} - \overline{L}_{-1}^{\text{[R]}}\big]\ket{B} = 0\\
&\qquad\qquad\implies \frac{\bra{\Omega}L_n^{[I]}\overline{L}_n^{[J]}\ket{B}}{\bra{\Omega}B\rangle} = \frac{n(n^2-1)}{6}R_{IJ}(\ket{B}).\label{recursivetransport}
\end{split}
\end{equation}
It is important to bear in mind that the formulas \eqref{transRefGen} are only one way to quantify transport probabilities, and they do so in an indirect and somewhat incomplete way because they only use Virasoro modes. Nonetheless, the boundary-state formalism had been extended by \cite{Kimura:2014hva} to include overlaps with states created by other symmetry generators besides those of Virasoro, but each associated matrix only gives one piece of CFT data.

\cite{Meineri:2019ycm} presents another more physically motivated and complete approach. Instead of folding the ICFT and invoking the boundary-state formalism, they set up a scattering \textit{gedankenexperiment} in the ICFT by sending a wave packet towards the interface and define transport coefficients in terms of ratios of average energy fluxes. The resulting ``scattering" transmission coefficients are generically determined by the coefficients of two-point functions between spin-2 quasi-primary holomorphic currents on the left and on the right of the interface.

Notably, if the only such current on both sides is the stress tensor, then the transmission coefficient is \textit{universal}---regardless of the in-state, transmission is a function of just one piece of CFT data $c_{\text{LR}}$ (keeping $c_{\text{L}},c_{\text{R}}$ fixed), and this quantity appears in the following two-point function of the left and right stress tensors computed in the vacuum state of the interface theory's Hilbert space:
\begin{equation}
\expval{T_{\text{L}}(z) T_{\text{R}}(z')}_{\text{interface}} = \frac{c_{\text{LR}}/2}{(z-z')^4}.\label{cldefinition}
\end{equation}
If transport only cares about $c_{\text{LR}}$, then the coefficients \eqref{transRefGen} defined previously by \cite{Quella:2006de} can be seen to describe \textit{weighted averages} of the scattering coefficients, as found by \cite{Meineri:2019ycm}. So \eqref{transRefGen} are sometimes called ``weighted-average coefficients" (e.g. by \cite{Bachas:2020yxv}).

In generic CFTs, the transport coefficients of \cite{Meineri:2019ycm} are state-dependent. Specifically, the in-state of the scattering experiment is created by some local operator, which \cite{Meineri:2019ycm} takes to be localized on the left and denotes as $O_{\text{L}}$, acting on the ICFT vacuum. The $O_{\text{L}} \times O_{\text{L}}$ OPE might contain a spin-2 holomorphic (quasi-)primary $\mathcal{O}_{\text{L}} \neq T_{\text{L}}$. For any such operator, there is a nontrivial two-point function $\expval{\mathcal{O}_{\text{L}}(z)T_{\text{R}}(z')}_{\text{interface}}$ with the same pole structure as \eqref{cldefinition}. The transport coefficients for the in-state created by $O_{\text{L}}$ thus depends not only on $c_{\text{LR}}$ but also the coefficients of these other two-point functions. So proper characterization of energy transport across the interface requires inclusion of this additional data.

We can extract this data from the boundary state $\ket{B}$ associated to the interface. For example, consider a \textit{primary}\footnote{Quasi-primaries would yields a nontrivial central term in the commutator with $L_n$, but this would not alter the qualitative discussion below.} spin-2 holomorphic current $\mathcal{O}$ whose Laurent modes are
\begin{equation}
\mathcal{O}_n \equiv \frac{1}{2\pi i}\oint dz\,z^{n+1} \mathcal{O}(z) \implies [L_n,\mathcal{O}_m] = (n-m)\mathcal{O}_{n+m}.
\end{equation}
We then define the following quantity in analogy to \eqref{clrdefBdry}:
\begin{equation}
\alpha^{\mathcal{O}}_{\text{LR}}(\ket{B}) = 2\frac{\bra{\Omega}\mathcal{O}_2^{[\text{L}]}\overline{L}_2^{[\text{R}]}\ket{B}}{\bra{\Omega}{B}\rangle}.\label{defalpha}
\end{equation}
Quantities defined with higher modes are redundant, again because of a recursion relation of the form \eqref{recursivetransport}. So $\alpha^{\mathcal{O}}_{\text{LR}}$ is the specific piece of CFT data associated with $\mathcal{O}$ that describes energy transport. However, we reiterate that if the scattering state is prepared with the stress tensor (i.e. $O_{\text{L}} = T_{\text{L}}$), then even in a generic CFT the only datum that matters in describing transport in that state is $c_{\text{LR}}$.

At this point, we focus on how the transport data of the symmetric orbifold is informed by that of the seed theory following the boundary state formalism of \cite{Quella:2006de,Kimura:2014hva,Kimura:2015nka}. We start with $c_{\text{LR}}$. We then discuss the data associated with another type of spin-2 holomorphic primary $W$ whose existence is rooted solely in the product structure of the symmetric orbifold theory. The punchline is that $\alpha^W_{\text{LR}} = 0$, so if transport in the seed theory only depends on $c_{\text{LR}}$, then the same is true in the symmetric orbifold.

\paragraph{Transmission of stress tensor}

Our first goal is to write the datum $c_{\text{LR}}$ for a symmetric orbifold theory's boundary state in terms of seed data. This will determine the $\mathcal{T},\mathcal{R}$ coefficients \eqref{transRefGen}.

First, recall that the $n$th full Virasoro generator is a sum over $n$th Virasoro generators of each copy of the seed theory. So, we have that
\begin{equation}
L_{2}^{[{I}]}\overline{L}_{2}^{[{J}]} = \left(\sum_{i=1}^N L_{2}^{(i)[{I}]}\right)\left(\sum_{j=1}^N \overline{L}_{2}^{(j)[{J}]}\right) = \sum_{i=1}^N L_{2}^{(i)[I]} \overline{L}_2^{(i)[J]} + \sum_{i \neq j} L_{2}^{(i)[I]} \overline{L}_2^{(j)[J]}.
\end{equation}
Observe that we have decomposed the sum into two pieces, where the second sum consists of ``cross terms." So to compute $R_{\text{LR}}$ (which is sufficient to get $\mathcal{T}$ and $\mathcal{R}$), we write
\begin{align}
R_{\text{LR}}(\ket{B})
&= \frac{\bra{0}^{\otimes N} L_2^{[\text{L}]} \overline{L}_2^{[\text{J}]} \ket{B}}{\bra{0}^{\otimes N}\ket{B}}\nonumber\\
&= \sum_{i=1}^N \frac{\bra{0}L_{2}^{(i)[\text{L}]} \overline{L}_2^{(i)[\text{R}]}\ket{b_{\alpha_i}}}{\bra{0}b_{\alpha_i}\rangle} + \sum_{i \neq j} \frac{\bra{0}L_{2}^{(i)[\text{L}]}\ket{b_{\alpha_i}}}{\bra{0}b_{\alpha_i}\rangle} \frac{\bra{0}\overline{L}_2^{(j)[\text{R}]}\ket{b_{\alpha_j}}}{\bra{0}b_{\alpha_j}\rangle}.
\end{align}
Here we have used the facts that the twisted-sector terms are projected out of the boundary state and that the normalization factors of $\ket{B}$ cancel. However, for $R_{\text{LR}}(\ket{B})$ to be purely a function of seed transport data $R_{\text{LR}}(\ket{b_{\alpha_i}})$, the cross terms above need to vanish. Indeed, we assert that these terms must vanish because their numerators are basically products of overlaps between seed conformal boundary states and spin-2 states. To see this explicitly, we invoke the gluing condition and the fact that global Virasoro generators annihilate $\bra{0}$:
\begin{align}
0 &= \bra{0}L_n^{(i)[I]} \left[L_{-1}^{(i)[\text{L}]} + L_{-1}^{(i)[\text{R}]} - \overline{L}_{1}^{(i)[\text{L}]} - \overline{L}_{1}^{(i)[\text{R}]}\right]\ket{b_{\alpha_i}}\nonumber\\
&= \bra{0}[L_{n}^{(i)[I]},L_{-1}^{(i)[I]}]\ket{b_{\alpha_i}}\nonumber\\
&= (n+1)\bra{0}L_{n-1}^{(i)[I]}\ket{b_{\alpha_i}}.
\end{align}
Thus, by taking $n = 3$ we have that $\bra{0}L_{2}^{(i)[I]}\ket{b_{\alpha_i}} = 0$, and so
\begin{equation}
R_{\text{LR}}(\ket{B}) = \sum_{i=1}^N R_{\text{LR}}(\ket{b_{\alpha_i}} \implies c_{\text{LR}}(\ket{B}) = \sum_{i=1}^N c_{\text{LR}}(\ket{b_{\alpha_i}}).\label{cLROrbi}
\end{equation}
Let us also write this in terms of the weighted-average transport coefficients \eqref{transRefGen}. First, we note that the seed transmission coefficient for a boundary state $\ket{b_{\alpha}}$ is
\begin{equation}
\mathcal{T}_{b_{\alpha}} \equiv \frac{2c_{\text{LR}}(\ket{b_\alpha})}{c_{\text{L}}^{(\text{s})} + c_{\text{R}}^{(\text{s})}},
\end{equation}
where $c_{\text{L}}^{(\text{s})}$ and $c_{\text{R}}^{(\text{s})}$ are respectively the central charges of the left and right seed theories. From \eqref{transRefGen}, we have that the transmission and reflection coefficients encoded by $\ket{B}$ are
\begin{align}
\mathcal{T}(\ket{B}) &= \frac{2 c_{\text{LR}}(\ket{B})}{Nc_{\text{L}}^{(\text{s})}+Nc_{\text{R}}^{(\text{s})}} = \frac{1}{N} \sum_{i=1}^N \mathcal{T}_{\alpha_i},\label{transTransOrbi}\\
\mathcal{R}(\ket{B}) &= \frac{2}{Nc_{\text{L}}^{(\text{s})}+Nc_{\text{R}}^{(\text{s})}} \sum_{i=1}^N \left[R_{\text{LL}}(\ket{b_{\alpha_i}}) + R_{\text{RR}}(\ket{b_{\alpha_i}})\right] = \frac{1}{N} \sum_{i=1}^N \mathcal{R}_{\alpha_i}.\label{transRefOrbi}
\end{align}
In other words, the weighted-average coefficients encoded by some boundary state $\ket{B}$ are themselves averages the coefficients encoded by each individual seed boundary state in $\ket{B}$.

\paragraph{Other spin-2 primary currents}

We have written $c_{\text{LR}}$ and the associated transport coefficients \eqref{transRefGen} in terms of seed data. However, there is in principle more data that goes into the energy transport coefficients, particularly if there are holomorphic quasi-primary currents of spin 2 apart from the stress tensor. It is indeed possible to have such currents in a symmetric orbifold, even when the seed theory does not. However, we argue that the associated transport data vanishes, and this intuitively is because the different seed copies do not interact at the orbifold point.

Let us be more specific about the construction of these spin-2 currents. It is a generic feature of symmetric orbifold theories that higher-spin (quasi-)primary currents can be written from those of the seed theory. Basically, such currents are constructed by starting with permutation-symmetric products of lower-spin (quasi-)primary currents and potentially adding derivative terms. We can check that the resulting quantity is itself primary by computing the OPE with the full stress tensor. The spin is easily seen to be the sum of the constituent factors' spins by applying a generic conformal transformation.

The existence of higher-spin currents is explored in depth by \cite{Apolo:2022fya}. An example found for all CFTs provided therein is the following holomorphic spin-4 primary current constructed solely from the seed stress tensor:
\begin{equation}
\sum_{i=1}^N \left[T^{(i)}T^{(i)} - \frac{3}{10}\partial^2 T^{(i)}\right] - \frac{22+5c}{5c(N-1)}\sum_{i \neq j} T^{(i)}T^{(j)}.
\end{equation}
Here, $T^{(i)}$ is the $i$th seed stress tensor, and as before $c$ is the seed central charge.

Note that these novel currents are always higher in spin relative to the seed currents. So, if we for example only have Virasoro symmetry in the seed theory, then there will be no spin-2 currents apart from the full stress tensor in the symmetric orbifold. However, if we have a seed current of spin $s = \frac{2}{k}$ for any positive integer $k \leq N$, then the symmetric orbifold should have a spin-2 current (built from $k$-fold products of seed operators), even if the seed theory does not!

To make the discussion even more specific, suppose that the seed theory has a U$(1)$ holomorphic current $J$ corresponding to some extended symmetry.\footnote{One could consider other symmetries, such as non-abelian ones. However, we do not anticipate the details of the group being relevant to the following statements.} \cite{Apolo:2022fya} constructs two spin-2 holomorphic fields:
\begin{equation}
Y \equiv \sum_{i=1}^N \left[T^{(i)} - \frac{3}{2}J^{(i)}J^{(i)}\right],\ \ W \equiv \sum_{i \neq j} J^{(i)}J^{(j)}.
\end{equation}
$Y$ is quasi-primary while $W$ is primary. Of course, our goal here is to consider symmetric orbifolds for which the $i$th seed theory does not have a spin-2 (quasi-)primary apart from $T^{(i)}$, so we can assume that $T^{(i)}$ is given by the Sugawara construction utilizing $J^{(i)}$ (as in a WZW model). In that case, $Y$ is actually just the full stress tensor (up to some rescaling), and so the only spin-2 holomorphic current present in the symmetric orbifold besides the full $T$ is $W$. This is enough to violate the assumptions in the proof of \cite{Meineri:2019ycm} that transport only cares about $c_{\text{LR}}$.

At this point, we seek to compute $\alpha_{\text{LR}}^{W}$ using the same boundary-state approach as for $c_{\text{LR}}$ above. To do so, first take the Laurent modes of $J^{(i)}$ and $W$:
\begin{equation}
J^{(i)}_n \equiv \frac{1}{2\pi i}\oint dz\,z^{n}J^{(i)}(z),\ \ W_n \equiv \frac{1}{2\pi i}\oint dz\,z^{n+1}W(z).
\end{equation}
Of course, each $W_n$ can be expressed as
\begin{equation}
W_n = \sum_{m \in \mathbb{Z}} \sum_{i \neq j} J^{(i)}_m J^{(j)}_{n-m}.\label{Wdefmodes}
\end{equation}
We now plug into \eqref{defalpha} to write
\begin{equation}
\alpha_{\text{LR}}^{W} = 2\sum_{i \neq j} \sum_{k=1}^N \frac{\bra{0}^{\otimes N}J_1^{(i)[\text{L}]} J_{1}^{(j)[\text{L}]} \overline{L}_2^{(k)[\text{R}]}\ket{B}}{\bra{0}^{\otimes N}\ket{B}}.
\end{equation}
When plugging in \eqref{Wdefmodes}, only the $m = 1$ term will survive acting on the vacuum, which is why we only have $J_1$ insertions. However, each term will contain a factor of the form $\bra{0}J_1^{(i)[I]}\ket{b_{\alpha_i}}$, which again is the overlap of a state with spin and a boundary state and will thus vanish by invoking the gluing condition \eqref{bcseedcond}.\footnote{To show this more explicitly, we can start with $\bra{0}J_n^{(i)[I]} \left[L_{-1}^{(i)[\text{L}]} + L_{-1}^{(i)[\text{R}]} - \overline{L}_{1}^{(i)[\text{L}]} - \overline{L}_{1}^{(i)[\text{R}]}\right]\ket{b_{\alpha_i}} = 0$ and invoke the commutator $[J_{n}^{(i)[I]},L_{-1}^{(i)[I]}] = nJ_{n-1}^{(i)[I]}$. Setting $n 
= 2$ yields $\bra{0}J_1^{(i)[I]}\ket{b_{\alpha_i}} = 0$.} So,
\begin{equation}
\alpha_{\text{LR}}^W = 0.\label{alphaWOrbi}
\end{equation}
Just to reiterate, we can consider other symmetry algebras. This would introduce nontrivial structure constants in the commutator between same-seed copies of currents. However, so long as we only have one spin-2 quasi-primary (i.e. the Sugawara stress tensor) in the seed theory, we still have that transport depends only on one parameter in the symmetric orbifold due to the non-interaction between different seed copies.

Indeed, all of the above equations and statements are only expected to hold at the orbifold point. In the strongly coupled regime far from the orbifold fixed point, mixing between the different copies would spoil \eqref{cLROrbi} and \eqref{alphaWOrbi}. One known way to access this sector is through holography, as we discuss in Section \ref{sec:t4Symm}.

\subsection{Extracting twisted-sector data}\label{sec:twistedData}

Symmetric orbifold theories have a large symmetry algebra that universally includes fractional Virasoro operators. These fractional modes can be used to define twisted-sector analogs to the untwisted data covered in Section \ref{sec:transportSeed}. We briefly elaborate on this point, but we leave much to follow-up work.

\paragraph{Twisted $g$-function and fractional entropy}

We start with the twisted $g$-function. Consider a twisted sector labeled by the integer partition $\{k_1,\dots,k_m\}$. By analogy to the untwisted quantity \eqref{gfunc}, we define the twisted $g$-function as the overlap of a boundary state $\ket{B}$ with the appropriate bare twist $\ket{0^{(k_1)}} \odot \cdots \odot \ket{0^{(k_m)}}$:
\begin{equation}
g^{(k_1,\dots,k_m)}(\ket{B}) \equiv \big(\bra{0^{(k_1)}} \odot \cdots \odot \bra{0^{(k_m)}}\big) \ket{B}.
\end{equation}
How do we interpret this quantity? First, we recall that the typical $g$-function is related to the boundary entropy $S_{\text{b}} = \log g$ that appears as a universal term in the thermodynamic limit of the entropy \cite{Cardy:1984bb,Cardy:2004hm}. This is often demonstrated by starting with the Euclidean cylinder partition function between two boundary states $\ket{A}$ and $\ket{B}$ in the closed-string quantization scheme. Following the normalization conventions of the review in \cite{Biswas:2022xfw}, we write this partition function as
\begin{equation}
Z = \bra{A}e^{-W H^{\text{cl}}} \ket{B},\label{zclosed}
\end{equation}
where $W$ is the width of the cylinder and $H^{\text{cl}}$ is the Hamiltonian that evolves the closed-string state on the circle from $\ket{A}$ to $\ket{B}$. In terms of (full) Virasoro modes and the circumference $\beta$, this is
\begin{equation}
H^{\text{cl}} = \frac{2\pi}{\beta}\left(L_0 + \overline{L}_0 - \frac{Nc}{12}\right),
\end{equation}
where we recall that the seed central charge is $c$, so the symmetric orbifold's is $Nc$.

We can identify \eqref{zclosed} with a thermal partition function (associated with the open-string quantization) and take the thermodynamic limit $W \gg \beta$. This is called the ``closed-string limit" and, assuming an Ishibashi basis $\{\kket{\mathfrak{h}}\}$, yields the approximation
\begin{equation}
Z \approx \sum_\mathfrak{h} \bra{A}{\mathfrak{h}}\rangle \bra{\mathfrak{h}}B\rangle q^{\mathfrak{h} - Nc/24},\ \ q \equiv e^{-4\pi W/\beta}.\label{closedstringPart}
\end{equation}
where we sum over primary states of weight $\mathfrak{h}$. The term that dominates in the strict limit is the one corresponding to the true vacuum $\mathfrak{h} = 0$, and so we can write the entropy as
\begin{equation}
S_{\text{th}} \equiv -\beta^2 \pdv{}{\beta} \big(\beta^{-1} \log Z\big) = \left(\frac{\pi N c}{3}\right) \frac{W}{\beta} + \log g(\ket{A}) + \log g(\ket{B}) + \cdots,
\end{equation}
i.e. as an extensive piece plus two constant pieces.

This limit produces an expression that is universal to all CFTs. However, in a symmetric orbifold theory, we can rearrange the partition function in the closed-string limit \eqref{closedstringPart} into an overall sum over twisted sectors, then note that each excited state within a particular twisted sector is suppressed by the bare twist.\footnote{This argument should be true for any theory with superselection sectors.} Thus we have the approximation
\begin{equation}
Z \approx \sum_{\substack{k_{1} + \cdots + k_m = N\\k_1 \leq \cdots \leq k_m}} g^{(k_1,\dots,k_m)}(\ket{A})\, g^{(k_1,\dots,k_m)}(\ket{B})\, q^{\mathfrak{h}[0^{(k_1)},\dots,0^{(k_m)}] - Nc/24},\label{partSymmorbi}
\end{equation}
in which twisted $g$-functions appear at subleading order to the vacuum term. We deduce that twisted $g$-functions represent subleading terms in the thermodynamic entropy.

\paragraph{Twisted transport matrices}

We now return to transport. For brevity and concreteness, we focus on the fractional Virasoro modes associated with the maximally twisted sector, which are $\{\ell_{n/N}^{\text{[L]}},\overline{\ell}_{n/N}^{\text{[L]}}\}$ for the theory left of the defect and $\{\ell_{n/N}^{\text{[R]}},\overline{\ell}_{n/N}^{\text{[R]}}\}$ for the theory right of the defect. However, the discussion here can be generalized to other modes. We also emphasize that we will continue using the transport-matrix formalism of \cite{Quella:2006de}. It remains to be seen how the quantities defined hereafter relate to CFT data probed through the more physical scattering processes studied by \cite{Meineri:2019ycm}.

The transport matrix \eqref{matDefGen} specifically computes transmission and reflection coefficients associated to the full stress tensor of the theory. One could instead construct a matrix that captures the physics of other fields, as in \cite{Kimura:2014hva,Kimura:2015nka}. To do so, we would need to employ the generators of the extended symmetry algebra. One possible candidate is
\begin{equation}
\widetilde{R}_{IJ}(\ket{B}) \equiv \frac{\bra{0}^{\otimes N} \ell^{[{I}]}_{n/N} \overline{\ell}^{[{J}]}_{n/N}\ket{B}}{\bra{0}^{\otimes N}\ket{B}},
\end{equation}
where as in \cite{Kimura:2014hva,Kimura:2015nka} we use the true vacuum of the theory $\ket{0}^{\otimes N}$. This would measure the transmission of some ``twisted" energy flux. Another possibility is to employ the maximally twisted bare twist $\ket{0^{(N)}}$, instead of the vacuum. By doing so, we define
\begin{equation}
R_{IJ}^{(N)}(\ket{B}) \equiv \frac{\bra{0^{(N)}} \ell_{2/N}^{[I]} \overline{\ell}_{2/N}^{[J]}\ket{B}}{\bra{0^{(N)}}B\rangle},
\end{equation}
in direct analogy to \eqref{matDefGen}. Note that we must use $\ell_{2/N}$ because $\bra{0^{(N)}}\ell_{1/N} = 0$ \cite{Burrington:2018upk}, much like how $\bra{0}L_1 = 0$ for any CFT with vacuum $\ket{0}$.

If the boundary state respects the extended algebra [meaning \eqref{maxTwConstraint} is satisfied], then we can derive a relation between matrices using higher fractional modes and $R_{IJ}^{(N)}$, just like \eqref{recursivetransport} for $R_{IJ}$. Specifically, we use the fact that
\begin{equation}
\bra{0^{(N)}}\ell_{n/N}^{[I]} \overline{\ell}_{(n+1)/N}^{[J]}\big[\ell_{1/N}^{\text{[L]}} + \ell_{1/N}^{\text{[R]}} - \overline{\ell}_{-1/N}^{\text{[L]}} - \overline{\ell}_{-1/N}^{\text{[R]}}\big] \ket{B} = 0,
\end{equation}
and the commutation bracket \eqref{fracVirModes} to write the recursive relation
\begin{equation}
\frac{n-1}{N} \bra{0^{(N)}}\ell_{(n+1)/N}^{[I]} \overline{\ell}_{(n+1)/N}^{[J]}\ket{B} = \frac{n+2}{N} \bra{0^{(N)}}\ell_{n/N}^{[I]} \overline{\ell}_{n/N}^{[J]}\ket{B}.
\end{equation}
After cancelling the $\frac{1}{N}$, we can see that the prefactors are identical to those of the recursive relation used to derive \eqref{recursivetransport}. Thus,
\begin{equation}
\frac{\bra{0^{(N)}} \ell_{n/N}^{[I]} \overline{\ell}_{n/N}^{[J]}\ket{B}}{\bra{0^{(N)}}B\rangle} = \frac{n(n^2 - 1)}{6}R_{IJ}^{(N)}(\ket{B}).
\end{equation}
Yet another matrix with perhaps with the clearest physical connection to scattering \cite{Meineri:2019ycm} is
\begin{equation}
R^{\ell}_{IJ}(\ket{B}) \equiv \frac{\bra{0^{(N)}} \ell^{[{I}]}_{2/N} \overline{L}^{[{J}]}_{2}\ket{B}}{\bra{0}^{\otimes N}\ket{B}},
\end{equation}
This should compute energy transmission of a scattering state prepared by exciting the bare twist $\ket{0^{(N)}}$ with the stress tensor.

We refrain from pushing these alternative matrices further for now. An interesting question is how to write these quantities in terms of the seed transport matrices. One path to do so might be to employ the covering-space techniques of \cite{Lunin:2000yv,Lunin:2001pw} used to study fractional modes by \cite{Burrington:2018upk,Burrington:2022dii,Burrington:2022rtr}. Another avenue is to directly scatter excitations of bare twists (since they are essentially twisted-sector vacua) and apply the definitions of \cite{Meineri:2019ycm}. We leave these directions to future work.

\section{The holographic symmetric orbifold of the $\mathbb{T}^4$ ICFT}\label{sec:t4Symm}

As a particularly tractable example, we examine the transport datum $c_{\text{LR}}$ in the symmetric orbifold of a $\mathbb{T}^4$ sigma model. The seed ICFT consists of four copies of a scalar field $\widetilde{\phi}:\mathbb{R} \to S^1$ which ``jumps" in coupling at some interface. To approximate the weakly coupled regime, we take the seed theory to consist of free scalars described by the action
\begin{equation}
I_{S^1} = \frac{R_-^2}{2} \int_{\tilde{y} < 0} dt\,d\tilde{y}\,(\partial \widetilde{\phi}_-)^2 + \frac{R_+^2}{2} \int_{\tilde{y} > 0} dt\,d\tilde{y}\,(\partial \widetilde{\phi}_+)^2.\label{freescalar}
\end{equation}
The couplings $R_{\pm}$ are the radii of the target space on the two sides of the interface and are generically distinct from one another.\footnote{Note that we are assuming the target space to be a square torus, with all of the radii equal.}

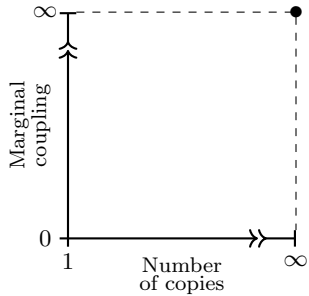
\begin{figure}
\centering
\begin{tikzpicture}
\draw[-,thin,dashed] (0.11,3) to (3,3) to (3,0.11);

\draw[->,thick] (0,0) to (0,2.5);
\draw[-,thick] (0,0) to (-0.11,0);
\draw[>-|,thick] (0,2.5) to (0,3);
\draw[->,thick] (0,0) to (2.5,0);
\draw[-,thick] (0,0) to (0,-0.11);
\draw[>-|,thick] (2.5,0) to (3,0);

\node[rotate=90] at (-0.5,3/2) {$\substack{\text{Marginal}\\\text{coupling}}$};
\node[] at (3/2,-0.5) {$\substack{\text{Number}\\\text{of copies}}$};
\node[rotate=90,white] at (3,3/2) {$\substack{\text{Marginal}\\\text{coupling}}$};

\node at (-0.3,0) {\footnotesize$0$};
\node at (-0.3,3) {\footnotesize$\infty$};
\node at (0,-0.3) {\footnotesize$1$};
\node at (3,-0.3) {\footnotesize$\infty$};

\node[] at (3,3) {$\bullet$};
\end{tikzpicture}

\caption{The parameter space of the symmetric orbifold of $\mathbb{T}^4$. We can dial both the number of seed copies and the marginal coupling which makes the copies interact. The bottom axis (with coupling $= 0$) describes the orbifold point. The black point is the holographic limit at which the description by type IIB supergravity on Janus is valid.}
\label{figs:janusICFTparams}
\end{figure}

We can turn on a marginal coupling that breaks the permutation symmetry and thus deforms the symmetric orbifold theory away from the orbifold point \cite{Avery:2010er}. The strongly coupled/deformed, large-$N$ regime of the symmetric orbifold theory built from scalars with the simple jump in \eqref{freescalar} is described by type IIB supergravity on 3-dimensional Janus \cite{Bak:2007jm} (Figure \ref{figs:janusICFTparams}). This background is a solution to the type IIB supergravity action constructed by deforming the $\text{AdS}_3 \times S^3 \times \mathbb{T}^4$ vacuum with a dilaton. 3d Janus is under a large amount of analytic control, just like the thin-brane models used in previous studies of holographic interfaces \cite{Bachas:2020yxv,Bachas:2021tnp,Anous:2022wqh,Baig:2022cnb}.

It is possible to compute $c_{\text{LR}}$ in the strongly coupled regime for the interface described by the Janus solution through recently developed methods \cite{Baig:2022cnb,Bachas:2022etu}. Additionally, working directly in pure Janus specifies a particular boundary state on the field-theory side constructed from copies of a single boundary state of the seed $S^1$ theory \cite{Azeyanagi:2007qj,Chiodaroli:2010ur}. With this in mind, we can also approximate the associated transport coefficients in the weakly coupled regime because the free-scalar seed theory \eqref{freescalar} is rather simple (cf. \cite{Quella:2006de}). Upon doing so, we compare the answer against that of \cite{Bachas:2022etu} to bound the running of transport coefficients with the marginal coupling of the ICFT. The resulting plot is shown in Figure \ref{figs:transComp}.

%%%%%%%%%%%%%%%%%%%%%%%%%%%%%%%
\subsection{3d Janus and holography}\label{sec:janusHolo}
%%%%%%%%%%%%%%%%%%%%%%%%%%%%%%%

\paragraph{Gravitational solution} In the Einstein frame of the type IIB supergravity action, we consider solutions of the form
\begin{equation}
ds^2_{\text{IIB}} = e^{\phi/2}\left(ds_{3}^2 + d\Omega_3^2\right)^2 + e^{-\phi/2} ds_{\mathbb{T}^4}^2.\label{sugraMetric}
\end{equation}
The dimensional reduction of the action onto the 3 noncompact dimensions is
\begin{equation}
S^{(3)}_{\text{GR}}[g,\phi] = \frac{1}{16\pi G_{\text{N}}^{(3)}} \int d^3 x\sqrt{-g}\left(R + \frac{2}{L^2} - \partial_\alpha \phi \partial^\alpha \phi\right),
\end{equation}
where $G_{\text{N}}^{(3)}$ is the 3d Newtonian constant, $L$ is a length scale, and Greek indices run over spacetime. The classical equations of motion are
\begin{align}
G_{\mu\nu} - \frac{1}{L^2}g_{\mu\nu} &= \partial_\mu \phi \partial_\nu\phi - \frac{g_{\mu\nu}}{2}\partial_\alpha \phi \partial^\alpha \phi,\\
\nabla_\mu \nabla^\mu \phi &= 0.
\end{align}
For the 3d spacetime, we assume the ansatz
\begin{equation}
\frac{ds^2}{L^2} = \frac{f(y)}{z^2}(-dt^2 + dz^2) + dy^2,\ \ \phi = \phi(y),\label{ansatz1}
\end{equation}
with $t,y \in (-\infty,\infty)$ and $z > 0$. The 3d Janus solution corresponds to the metric function,
\begin{equation}
f(y) = \frac{1}{2}\left[1+\sqrt{1-2\gamma^2}\cosh(2y)\right],
\end{equation}
where $\gamma \in \left[0,\frac{1}{\sqrt{2}}\right)$ is some parameter characterizing the solution.\footnote{$\gamma = 0$ corresponds to the non-deformed AdS$_3$ vacuum, and we find naked singularities if $\gamma^2 > \frac{1}{2}$ \cite{Bak:2007jm}.} See Figure \ref{figs:janus3} for a visual representation of 3d Janus. In this background, the dilaton's profile is
\begin{figure}
\centering
\begin{tikzpicture}[yscale=0.7]
\draw[<-,blue,thick] (-3,0) to (0,0);
\draw[->,red,thick] (0,0) to (3,0);

\draw[-,thick,dashed,red!17!blue!83] (0,0) to (-2.598,1.5);
\draw[-,thick,dashed,red!33!blue!67] (0,0) to (-1.5,2.598);
\draw[-,thick,dashed,blue!50!red!50] (0,0) to (0,3);
\draw[-,thick,dashed,blue!33!red!67] (0,0) to (1.5,2.598);
\draw[-,thick,dashed,blue!17!red!83] (0,0) to (2.598,1.5);

\node[scale=1.5,purple!60!blue] at (0,0) {$\bullet$};

\node[blue] at (-2.7,-0.5) {$y = -\infty$};
\node[red] at (2.7,-0.5) {$y = +\infty$};

\node[purple!60!blue] at (0,-0.5) {$y = 0$};

\draw[-,thick] (0,1.5) arc (90:105:1.5);
\draw[->,thick] (0,1.5) arc (90:75:1.5);

\draw[-,thick] (0,1.5) to (0,1.25);
\draw[->,thick] (0,1.5) to (0,2);
\node at (-0.25,2) {$z$};
\node at (0.5,1.7) {$y$};
\end{tikzpicture}
\caption{A constant-time slice of $3$-dimensional Janus with radial coordinate $z$ and hyperbolic angular coordinate $y$. The dashed lines are AdS$_2$ slices of the bulk. The $y = 0$ point is the interface between the left ($y < 0$) and right ($y > 0$) CFTs.}
\label{figs:janus3}
\end{figure}
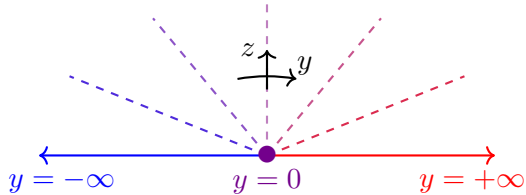
\begin{equation}
\phi(y) = \phi_0 + \frac{1}{\sqrt{2}}\log\left(\frac{1 + \sqrt{1-2\gamma^2}+ \sqrt{2}\gamma\tanh y}{1 + \sqrt{1-2\gamma^2}- \sqrt{2}\gamma\tanh y}\right),\label{dilProf}
\end{equation}
Thus, the asymptotic values of the dilaton are
\begin{equation}
\phi_{\pm} = \lim_{y \to \pm\infty} \phi(y) = \phi_0 \pm \frac{1}{2\sqrt{2}}\log\left(\frac{1+\sqrt{2}\gamma}{1-\sqrt{2}\gamma}\right).\label{asympDil}
\end{equation}
One benefit of studying the 3d Janus solution is that the dual field theory is known to be a highly deformed (i.e. strongly coupled) symmetric orbifold ICFT. The parameters of the theory can be described in terms of the number of fundamental D-branes.

\paragraph{Field-theory dual} The starting point is to recall the holographic dual of type IIB string theory on $\text{AdS}_3 \times S^3 \times \mathbb{T}^4$. There, we take the D1/D5 system consisting of $Q_1$ D1-branes and $Q_5$ D5-branes in $\mathbb{R}^6 \times \mathbb{T}^4$ \cite{Horowitz:1996ay}. For large $Q_1,Q_5 \gg \frac{1}{g_{\text{s}}} \gg 1$, the AdS/CFT correspondence describes a duality between type IIB supergravity on $\text{AdS}_3 \times S^3 \times \mathbb{T}^4$ and a strongly coupled CFT on the boundary of $\text{AdS}_3 \times S^3$. The target space of the CFT is topologically $(\mathbb{T}^4)^{Q_1 Q_5}/S_{Q_1 Q_5}$ \cite{Maldacena:1997re,Seiberg:1999xz,David:2002wn}, and so it is a highly deformed symmetric orbifold theory consisting of $N \equiv Q_1 Q_5 \gg 1$ copies of a $\mathbb{T}^4$ sigma model.

To construct the Janus solution, we are adding a (non-supersymmetric) dilatonic deformation on top of the AdS$_3 \times S^3 \times \mathbb{T}^4$ vacuum. As this dilaton asymptotes to two distinct values on the AdS boundary \eqref{asympDil}, the bulk deformation introduces an interface in the dual symmetric orbifold theory.

We can identify the radius of each $S^1$ factor of the target-space torus with the asymptotic value of $e^{-\phi/2}$ \cite{Azeyanagi:2007qj}. However, since the dilaton profile \eqref{dilProf} takes two values at infinity \eqref{asympDil}, the radius ``jumps" between two values $R_+$ and $R_-$, where
\begin{equation}
R_{\pm} \propto e^{-\phi_{\pm}/2} = e^{-\phi_0/2} \left(\frac{1+\sqrt{2}\gamma}{1-\sqrt{2}\gamma}\right)^{\mp \frac{1}{4\sqrt{2}}}.
\end{equation}
We interpret this nontrivial jump as describing the presence of an interface in the theory. With this in mind, observe that the ratio of the two radii,
\begin{equation}
\frac{R_+}{R_-} = \left(\frac{1+\sqrt{2}\gamma}{1-\sqrt{2}\gamma}\right)^{-\frac{1}{2\sqrt{2}}},\label{ratioJanusParam}
\end{equation}
goes to $1$ as $\gamma \to 0$ (the interface disappears) and goes to $0$ as $\gamma \to \frac{1}{\sqrt{2}}$ (there is a parametrically large separation of scales between the sides). Intuitively, we expect these limits to respectively correspond to a completely transparent or completely reflective interface.

Note that the form of the boundary state encoding the Janus interface is not obvious. However, \cite{Azeyanagi:2007qj} argues that it consists of $N$ identical copies of a four-fold (one per $S^1 \subset \mathbb{T}^4$) ``Neumann--Dirichlet" state in the $\mathbb{T}^4$ sigma model. While there may be twisted-sector terms, from our earlier analysis we assert that transport is not sensitive to the resulting combinatorial factors and does not probe such terms.

\subsection{Strong-coupling transport from gravity waves}

The holographic prescription for the transport datum $c_{\text{LR}}$ \cite{Bachas:2020yxv} is to consider linearized, source-free fluctuations of the metric in Fefferman--Graham (FG) gauge \cite{Skenderis:1999nb} (setting $L = 1$),
\begin{equation}
ds^2_{\text{FG}} = \frac{du^2}{u^2} + \frac{1}{u^2}\left[g_{ij}^{(0)} + u^2 g_{ij}^{(2)} + \frac{u^4}{4}g_{ij}^{(4)} + \cdots\right]dw^i dw^j,\label{fgexp}
\end{equation}
where $u > 0$ is the radial coordinate and the Latin indices run over the remaining $1+1$ dimensions. These fluctuations are called \textit{surface gravity waves}, and in the boundary theory they corresponds to excitations produced by the stress tensor \cite{Skenderis:1999nb,Skenderis:2000in}. In the thin-brane model of \cite{Bachas:2020yxv} consisting of two AdS$_3$ geometries glued along an AdS$_2$ surface, we scatter these surface gravity waves off of the brane. This corresponds to the scattering \textit{gedankenexperiment} of \cite{Meineri:2019ycm} in the boundary theory, with the in-state being prepared by the stress tensor. The amplitudes of the reflected and transmitted waves can be translated into the transport coefficients. The bulk equations of motion and boundary conditions can then be used to constrain these transport coefficients.

\paragraph{Scattering on Janus}

We may attempt an analogous scattering experiment directly in Janus. While possible in principle, this is difficult in practice. To see why, we first write \eqref{ansatz1} in FG form to serve as the background of our scattering experiment. This has been done perturbatively in the Janus parameter by \cite{Papadimitriou:2004rz} (see also \cite{Estes:2014hka,Gutperle:2016gfe}). Up to the first subleading term of order $\gamma^2$, we approximate the metric function $f$ in \eqref{ansatz1} as
\begin{equation}
f(y) = \cosh^2 y - \frac{1}{2}\gamma^2 \cosh^2 y + O(\gamma^4).
\end{equation}
Taking $\gamma = 0$ (pure AdS), the FG metric comes about through the coordinate transformation $z \to \sqrt{\tilde{y}^2 + u^2}$ and $y \to \Sinh^{-1}\left(\frac{\tilde{y}}{u}\right)$. For Janus, we thus consider the ansatz
\begin{equation}
z \to \sqrt{\tilde{y}^2 + u^2} + \gamma^2 u f_z\left(\frac{\tilde{y}}{u}\right) + O(\gamma^4),\ \ y \to \Sinh^{-1}\left(\frac{\tilde{y}}{u}\right) + \gamma^2 f_y\left(\frac{\tilde{y}}{u}\right) + O(\gamma^4).\label{metTransFG}
\end{equation}
Upon plugging this in and insisting that the geometry is still asymptotically AdS (as $u \to 0$), we find the functions
\begin{align}
f_z(x) = \frac{-2 + (1+2x^2) \log\left(\frac{1+x^2}{x^2}\right)}{8\sqrt{1+x^2}},\ \ f_y(x) = \frac{1+4x^2 + x^2 \log\left(\frac{1+x^2}{x^2}\right)}{8x\sqrt{1+x^2}},
\end{align}
for which the transformed metric truncated at order $\gamma^2$ is\footnote{The transformation \eqref{metTransFG} induces terms which are $O(\gamma^4)$, but we omit these terms.}
\begin{equation}
\begin{split}
ds^2
&\sim \frac{du^2}{u^2} - \left[\frac{1}{u^2} + \frac{\gamma^2}{4}\left(\frac{1}{u^2 + \tilde{y}^2} - \frac{1}{u^2}\log\left(1 + \frac{u^2}{\tilde{y}^2}\right)\right)\right]dt^2\\
&\quad+ \left[\frac{1}{u^2} - \frac{\gamma^2}{4}\left(\frac{1}{\tilde{y}^2} - \frac{1}{u^2}\log\left(1 + \frac{u^2}{\tilde{y}^2}\right)\right)\right]d\tilde{y}^2\\
&= \frac{du^2}{u^2} + \frac{1}{u^2}\left[\left(-dt^2 + d\tilde{y}^2\right) - \frac{u^4}{4}\frac{\gamma^2}{2\tilde{y}^4} \left(-dt^2 + d\tilde{y}^2\right) + \cdots\right].
\end{split}\label{fgexp2}
\end{equation}
Unlike the thin-brane solution on AdS$_3$ of \cite{Bachas:2020yxv}, the metric does not truncate at the $g_{ij}^{(4)}$ term and breaks down near the interface. These features make it difficult to compute constraints from fluctuations on Janus even up to order-$\gamma^2$ terms. Additionally, the fluctuations may themselves need $\gamma$-dependence in order for the constraints to be nontrivial.

Most importantly, our goal is to perform a strong-weak comparison over the full range of the Janus parameter $\gamma$. As such, while a perturbative calculation might be a useful proof of principle, it is not particularly helpful in accomplishing our ultimate purpose.

\paragraph{Stacking branes}

\begin{figure}
\centering
\begin{tikzpicture}
\draw[->,blue,thick] (-4.5,0) to (-7.5,0);
\draw[->,red,thick] (4.5,0) to (7.5,0);

\draw[-,thick,dashed,red!17!blue!83] (-4.5,0) to (-3.5,2);
\draw[-,thick,dashed,red!17!blue!83] (-2,0) to (-3,2);
\draw[-,thick,dashed,red!33!blue!67] (-2,0) to (-1.25,2);

\draw[-,thick,dashed,red!33!blue!67] (0,0) to (-0.75,2);
\draw[-,thick,dashed,blue!33!red!67] (0,0) to (0.75,2);

\draw[-,thick,dashed,blue!17!red!83] (4.5,0) to (3.5,2);
\draw[-,thick,dashed,blue!17!red!83] (2,0) to (3,2);
\draw[-,thick,dashed,blue!33!red!67] (2,0) to (1.25,2);

\node[scale=1.5,purple!60!blue] at (-4.5,0) {$\bullet$};
\node[scale=1.5,purple!60!blue] at (4.5,0) {$\bullet$};

\node[scale=1.5,white] at (-2,0) {$\bullet$};
\node[scale=1.5,purple!60!blue] at (-2,0) {$\circ$};

\node[scale=1.5,white] at (2,0) {$\bullet$};
\node[scale=1.5,purple!60!blue] at (2,0) {$\circ$};

\node[scale=1.5,white] at (0,0) {$\bullet$};
\node[scale=1.5,purple!60!blue] at (0,0) {$\circ$};
\end{tikzpicture}
\caption{A discrete array of thin branes in AdS$_3$. This is the setup used by \cite{Baig:2022cnb} to extend the prescription for transport coefficients of \cite{Bachas:2020yxv} beyond the case of a thin brane with one tension parameter. ``Thick brane" configurations in which the interface is represented by a smooth geometry foliated into AdS slices can be discretized into an array of thin branes.}
\label{figs:arrayOfBranes}
\end{figure}
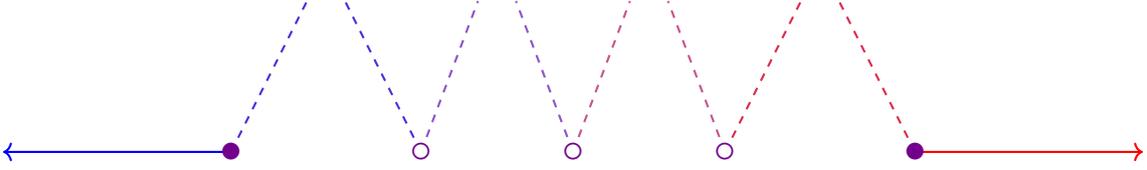

Fortunately, a different $\gamma$-exact calculation of the transport coefficients in 3d Janus has been performed by \cite{Bachas:2022etu}. The first step had been taken by \cite{Baig:2022cnb}, which extended the earlier thin-brane method to ``arrays" of thin branes (Figure \ref{figs:arrayOfBranes}). This provides more parametric freedom in the bulk and allows for holographic computations of transport coefficients encoded by a broader class of interfaces.

\cite{Bachas:2022etu} considers interfaces which are holographically described by continuous $(d+1)$-dimensional bulk geometries that can be foliated into AdS$_d$ slices, such as the Janus solution. They observe that these ``thick-brane" geometries can be treated as a limit of a discrete array of thin branes. For 3d Janus\footnote{\cite{Bachas:2022etu} uses a slightly different Janus parameter. We present their result in terms of $\gamma$.} in particular, they find a tension ``density" for the branes in this array and integrate to compute an ``effective" tension. Plugging this into the original thin-brane formula \cite{Bachas:2020yxv} yields a $\gamma$-exact holographic transmission coefficient:
\begin{equation}
\mathcal{T}^{\text{GR}} \equiv c_{\text{LR}}^{\text{Janus}} = \frac{\sqrt{2}\gamma}{\Tanh^{-1}(\sqrt{2}\gamma)} = 1 - \frac{2}{3}\gamma^2 - \frac{16}{45}\gamma^4 + O(\gamma^6).\label{transBC}
\end{equation}
This is expected to be equivalent to the answer obtained by scattering surface gravity waves directly on Janus. Nonetheless, the stacking approach is much less tedious and more powerful.

It is worth comparing the Janus result against that of the thin-brane models \cite{Bachas:2020yxv}. In the latter, the underlying action is Einstein plus a Randall--Sundrum term \cite{Randall:1999vf,Karch:2000ct}. The tension is thus an effective coupling constant, and tuning the transport coefficients requires fixing the tensions by hand---a fine-tuning problem. However, 3d Janus is a genuine top-down solution, and so we should not require fine-tuning to achieve a particular $\mathcal{T}$ and $\mathcal{R}$. Indeed, the Janus parameter $\gamma$ is an integration constant labeling the solutions rather than a coupling, and the full physical range of $\gamma$ furnishes the full unitary range $\mathcal{T} \in [0,1]$.

\subsection{Weak-coupling transport from seed theory}

We now calculate the transport coefficients in the free sector of the $\mathbb{T}^4$ symmetric orbifold ICFT. As we are at the orbifold point, we only need the transport coefficients of the individual seed boundary states used to construct the Janus interface. Then, just as in \cite{Azeyanagi:2007qj}, we take $4N = 4Q_1 Q_5$ copies of the Neumann--Dirichlet state in the free scalar theory. With the seed coefficients in hand, we then employ the prescription of Section \ref{sec:symmOrbi}---namely \eqref{transTransOrbi}--\eqref{transRefOrbi}---to write the ``full" transport coefficients.

The seed theory consists of four copies of a free scalar field on 2d Minkowski spacetime $-dt^2 + d\tilde{y}^2$ whose target space is $S^1$ and with an interface at $\tilde{y} = 0$:
\begin{equation}
S_{\text{FT}}[\widetilde{\phi}] = \frac{R_-^2}{2} \int_{\tilde{y} < 0} dt\,d\tilde{y}\,\partial_i \widetilde{\phi}_- \partial^i \widetilde{\phi}_- + \frac{R_+^2}{2} \int_{\tilde{y} > 0} dt\,d\tilde{y}\,\partial_i \widetilde{\phi}_+ \partial^i \widetilde{\phi}_+,
\end{equation}
where $i$ here is a spacetime index running over $(t,\tilde{y})$.

The transmission and reflection coefficients in this $S^1$ theory have been computed by \cite{Quella:2006de} (based on \cite{Bachas:2001vj}) in a different parameterization. For now, we simply need to recast their results in terms of $R_{\pm}$. To complete this exercise, we require the behavior of the fields at the interface. Reintroducing coordinate dependence as $\widetilde{\phi} \to \widetilde{\phi}(t,\tilde{y})$, we demand $\delta \widetilde{\phi}_+(t,0) = \delta \widetilde{\phi}_-(t,0)$ \cite{Azeyanagi:2007qj}. By varying the action and integrating by parts, we then get the boundary condition at $\tilde{y} = 0$:
\begin{equation}
R_+^2 \partial_{\tilde{y}} \widetilde{\phi}_+ = R_-^2 \partial_{\tilde{y}} \widetilde{\phi}_-.
\end{equation}
This can be rewritten as a matrix equation in the form presented by \cite{Quella:2006de}. Upon making the substitution $\pm\partial_{\tilde{y}} \to \partial_\pm$, we write 
\begin{equation}
\begin{pmatrix}
\partial_- \widetilde{\phi}_-\\
\partial_+ \widetilde{\phi}_+\\
\end{pmatrix} = S\begin{pmatrix}
\partial_+ \widetilde{\phi}_-\\
\partial_- \widetilde{\phi}_+\\
\end{pmatrix},\ \ S = \begin{pmatrix}
-\cos(2\theta) & \sin(2\theta)\\
\sin(2\theta) & \cos(2\theta)
\end{pmatrix},
\end{equation}
where we have identified the $\theta$ parameter of \cite{Quella:2006de} with $R_{\pm}$ as follows:
\begin{equation}
\cos(2\theta) = 1 - \dfrac{2R_-^4}{R_+^4 + R_-^4},\ \ \sin(2\theta) = \frac{2R_+^2 R_-^2}{R_+^4 + R_-^4}.
\end{equation}
The transmission and reflection coefficients in the free theory on $S^1$ are then
\begin{align}
\mathcal{T}_{S^1}^{\text{FT}} &= \sin^2(2\theta) = \frac{4}{\left[\left(\frac{R_+}{R_-}\right)^2 + \left(\frac{R_-}{R_+}\right)^2\right]^2},\label{transS1}\\
\mathcal{R}_{S_1}^{\text{FT}} &= \cos^2(2\theta) = \left[\frac{\left(\frac{R_+}{R_-}\right)^2 - \left(\frac{R_-}{R_+}\right)^{2}}{\left(\frac{R_+}{R_-}\right)^2 + \left(\frac{R_-}{R_+}\right)^{2}}\right]^2.\label{refS1}
\end{align}
The seed $\mathbb{T}^4$ sigma model consists of four non-interacting $S^1$ factors. For boundary states in the product theory, a similar argument to that of a symmetric orbifold theory at the orbifold point (Section \ref{sec:transportSeed}) applies; the transport coefficients of a product state are averages of the transport coefficients of the individual factors \eqref{transTransOrbi}--\eqref{transRefOrbi}. As the $S^1$ factors have the same boundary condition, we deduce that the transmission and reflection coefficients in the $\mathbb{T}^4$ theory are still given by \eqref{transS1}--\eqref{refS1}. Furthermore, as all of the seed boundary states are also identical, the total transmission and reflection coefficients (respectively $\mathcal{T}^{\text{FT}}$ and $\mathcal{R}^{\text{FT}}$) are simply the seed values, because they too are computed as averages.

Now, we may recast the transport coefficients in terms of the Janus parameter $\gamma$ using \eqref{ratioJanusParam}. Doing so for the transmission coefficient $\mathcal{T}^{\text{FT}}$, we have that
\begin{equation}
\mathcal{T}^{\text{FT}} \equiv c_{\text{LR}}^{\text{Sym}(\mathbb{T}^4)} = \frac{4\left(1+\sqrt{2}\gamma\right)^{\sqrt{2}}\left(1-\sqrt{2}\gamma\right)^{\sqrt{2}}}{\left[\left(1+\sqrt{2}\gamma\right)^{\sqrt{2}} + \left(1-\sqrt{2}\gamma\right)^{\sqrt{2}}\right]^2} = 1 - 4\gamma^2 + \frac{16}{3}\gamma^4 + O(\gamma^6).\label{transWeak}
\end{equation}

\subsection{Comparing across regimes}

\begin{figure}
\centering
\includegraphics[scale=0.835]{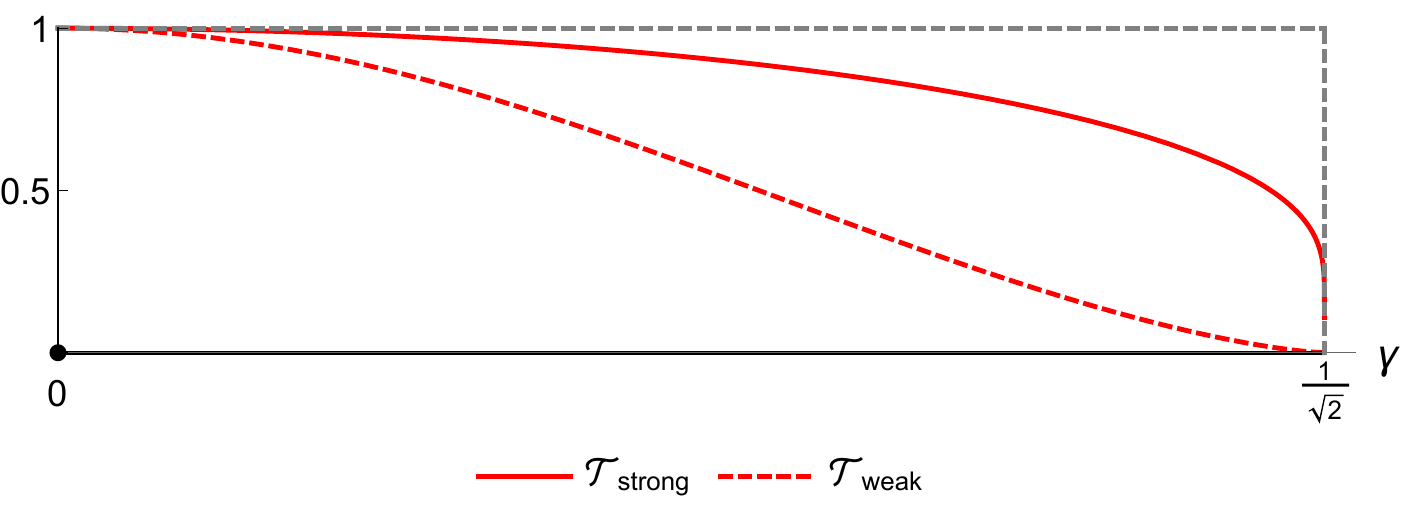}
\caption{The transmission coefficients as functions of the Janus parameter $\gamma \in \left[0,\frac{1}{\sqrt{2}}\right)$ both at strong coupling (solid) and at weak coupling (dashed). While they match at the extremal values of $\gamma$, the values at strong coupling are consistently larger than those at weak coupling. Furthermore, the weak-coupling coefficient has an inflection point at $\gamma \approx 0.406$, whereas the strong-coupling coefficient has a strictly negative derivative.}
\label{figs:transComp}
\end{figure}

Equipped with the transport coefficients of the Janus interface both at strong and weak coupling, we now compare the results. To reiterate, the strong-coupling transmission coefficient is approximately the answer obtained from the Janus solution,
\begin{equation}
\mathcal{T}_{\text{strong}} \sim \frac{\sqrt{2}\gamma}{\Tanh^{-1}(\sqrt{2}\gamma)} = 1 - \frac{2}{3}\gamma^2 - \frac{16}{45}\gamma^4 + O(\gamma^6),\label{strongT}
\end{equation}
while the weak-coupling transmission coefficient is approximately the one computed directly from the $\mathbb{T}^4$ symmetric orbifold theory at the orbifold point,
\begin{equation}
\mathcal{T}_{\text{weak}} \sim \frac{4\left(1+\sqrt{2}\gamma\right)^{\sqrt{2}}\left(1-\sqrt{2}\gamma\right)^{\sqrt{2}}}{\left[\left(1+\sqrt{2}\gamma\right)^{\sqrt{2}} + \left(1-\sqrt{2}\gamma\right)^{\sqrt{2}}\right]^2} = 1 - 4\gamma^2 + \frac{16}{3}\gamma^4 + O(\gamma^6).\label{weakT}
\end{equation}
These are both plotted in Figure \ref{figs:transComp}.

We immediately observe that $\mathcal{T}_{\text{strong}} > \mathcal{T}_{\text{weak}}$ away from the extremal values of $\gamma$. In other words, turning on the marginal coupling which deforms the symmetric orbifold theory also increases the proportion of energy transmitted through the interface at fixed $\gamma$. This makes intuitive sense---energy is able to be exchanged between different copies of the seed left and right CFTs, whereas at the orbifold point these copies do not interact at all.

We also observe that the two functions are structurally different. While $\mathcal{T}_{\text{strong}}(\gamma)$ has a strictly negative derivative and approaches $0$ rapidly, $\mathcal{T}_{\text{weak}}(\gamma)$ has an inflection point at $\gamma \approx 0.406$ and approaches $0$ more slowly. This indicates that the functional form of the coefficient $\mathcal{T}$ changes as the coupling runs.

We conclude by emphasizing that this result is very different from the situation for boundary entropy $S_{\text{b}}$ \cite{Azeyanagi:2007qj}, in which the analogous strong-weak comparison using the Janus solution involves two numerically similar functions of $\gamma$. The answer at strong coupling was approximated by employing the Ryu--Takayanagi formula:
\begin{equation}
\frac{S_{\text{b}}^{\text{strong}}}{N} \sim \log\left(\frac{1}{\sqrt{1-2\gamma^2}}\right) = \gamma^2 + \gamma^4 + O(\gamma^6),
\end{equation}
while the answer at weak coupling was found in field theory to be
\begin{equation}
\frac{S_{\text{b}}^{\text{weak}}}{N} \sim \log\left[\frac{\big((1+\sqrt{2}\gamma)^{1/\sqrt{2}} + (1-\sqrt{2}\gamma)^{1/\sqrt{2}}\big)^2}{4(1-2\gamma^2)^{1/\sqrt{2}}}\right] = \gamma^2 + \frac{7}{6}\gamma^4 + O(\gamma^6).
\end{equation}
Unlike the strong and weak values of $\mathcal{T}$ \eqref{strongT}--\eqref{weakT}, the strong and weak values of boundary entropy are the same at order-$\gamma^2$, and the difference at order-$\gamma^4$ is much smaller [an $O(1)$ fraction]. In other words, the boundary entropy appears to be more protected from the running of the coupling than transmission.

\section{Discussion}\label{sec:disc}

To summarize, we have first explored transport coefficients of interfaces in generic symmetric orbifold theories (taken at their orbifold points). In particular, we have used BCFT techniques to write them in terms of ``seed" transport coefficients, using the boundary-state construction of \cite{Belin:2021nck} and applying the transport-matrix approach of \cite{Quella:2006de}. We have found that, regardless of the number of copies $N$, the transport coefficients of the boundary states in a symmetric orbifold theory are averages of transport coefficients encoded by seed-theory boundary states, as per the formulas \eqref{transTransOrbi}--\eqref{transRefOrbi}.

The second part of this paper is a study of the $\mathbb{T}^4$ symmetric orbifold theory. This theory can be understood at strong marginal coupling (away from the orbifold point) through the AdS/CFT correspondence. A simple class of interfaces in this theory are described by the 3d Janus solution to type IIB supergravity \cite{Bak:2007jm}. From the tools of gravity, one can extract transport coefficients for this class of interfaces at strong coupling \cite{Bachas:2022etu}. Furthermore, we compute the transport coefficients at weak coupling (at the orbifold point) by combining our earlier methods and with our knowledge of the seed $\mathbb{T}^4$ sigma model.

This sets the stage for a comparison between the transport coefficients at strong coupling and at weak coupling for Janus interfaces in the symmetric orbifold of $\mathbb{T}^4$. We ultimately find a marked difference. The coefficients are structurally different functions of $\gamma$, but transmission through the interface is larger at strong coupling than at weak coupling.

\subsection{Transport versus thermodynamics}\label{sec:transport}

We reiterate that our final result in the symmetric orbifold of $\mathbb{T}^4$ is notably different from previous analogous computations of the boundary entropy \cite{Azeyanagi:2007qj,Chiodaroli:2010ur}. In those cases, boundary entropy of the Janus interface had been found to be relatively protected from the running of the coupling, with supersymmetry \textit{completely} protecting it \cite{Chiodaroli:2010ur}. Meanwhile, the transport coefficients develop different functional features entirely---most notably the loss of the inflection point at strong coupling in Figure \ref{figs:transComp}.

This is reasonable in light of other strong-weak comparisons in holography. One can look to the case of $\mathcal{N} = 4$ 4d supersymmetric Yang--Mills (SYM) theory, which at strong 't Hooft coupling is dual to type IIB supergravity on AdS$_5 \times S^5$. There, the free energy, which like boundary entropy is a thermodynamic quantity, has been computed both in the free theory and at strong coupling through holography \cite{Gubser:1998nz,Fotopoulos:1998es}. While the coupling runs over an infinite range, the free energy only changes by a finite factor of $\frac{3}{4}$.

Another quantity which has been compared at both regimes is the shear viscosity of the SYM plasma \cite{Policastro:2001yc,Buchel:2004di,Huot:2006ys}. This quantity is associated with transport, and its story is very different from that of the free energy. In units of entropy density, it is well-known that the shear viscocity reaches a finite value of $\frac{1}{4\pi}$ at strong coupling. However, it blows up at weak coupling. Thus, the functional dependence on coupling is not described by a finite interpolating function, unlike for free energy.

Of course, we are comparing different quantities---the boundary entropy versus the transmission coefficient---from those of the $\mathcal{N} = 4$ SYM story. However, they are still respectively facets of thermodynamics and transport, and we again see that the thermodynamic quantity (boundary entropy) is much more strongly \cite{Azeyanagi:2007qj} (or even completely \cite{Chiodaroli:2010ur}) protected from the running of the coupling than the transport quantity (transmission coefficient). If would thus be interesting to further scrutinize the validity of this idea that ``transport rushes as thermodynamics dawdles" in coupling, with Janus setups (including the higher-dimensional version \cite{Bak:2003jk,Clark:2005te}) being a realm for doing so.

\subsection{The Janus boundary state}\label{sec:janInt}

We emphasize that the specific form of the boundary state encoding a Janus interface is not obvious. To glean some insight, we look to the holographic calculation of the boundary entropy in \cite{Azeyanagi:2007qj} and assume that this should be (reasonably) protected from the running of the marginal coupling. Employing the Ryu--Takayanagi formula \cite{Ryu:2006bv} yields
\begin{equation}
S_{\text{b}}^{\text{GR}} = N \log\left(\frac{1}{\sqrt{1-2\gamma^2}}\right),\ \ N = Q_1 Q_5 \gg 1.\label{bdryEntropyRT}
\end{equation}
From this, we see that boundary entropy is an order-$N$ quantity at large $N$. Furthermore, in performing their comparison with the value at the orbifold point, \cite{Azeyanagi:2007qj} argues and uses the fact that all copies of the seed boundary state are identical---they are 4-fold products of a ``Neumann-Dirichlet" boundary state in the $S^1$ theory $\ket{b_{\text{ND}}}$.

However, this is not enough to specify the form of the boundary state $\ket{B_{\text{J}}}$. For example, we can imagine that it takes the form\footnote{This would not be the sort of a boundary state constructed by \cite{Belin:2021nck}.}
\begin{equation}
\ket{B_{\text{J}}} = \left(\ket{b_{\text{ND}}}^{\otimes 4}\right)^{\otimes N}.\label{bdryFactor}
\end{equation}
This is assumed by \cite{Azeyanagi:2007qj,Chiodaroli:2010ur} in computing boundary entropy at the orbifold point. In support of this, a lack of twisted-sector terms is not unreasonable. Janus is a type IIB supergravity vacuum. The density of states (in scaling dimension) of the ICFT dual to supergravity on Janus should obey supergravity-like (slow) growth (cf. \cite{Belin:2020nmp,Benjamin:2022jin,Apolo:2022fya}). Twisted sectors in the symmetric orbifold theory, however, obey Hagedorn (fast) growth at large $N$ \cite{Belin:2019rba}. As such, we might only expect to get an interface consistent with the Janus solution if we omit twisted-sector terms even in the free symmetric orbifold. Furthermore, that the two calculations in \cite{Azeyanagi:2007qj} match to a reasonable degree can be taken as additional evidence for the veracity of \eqref{bdryFactor} if one assumes that boundary entropy should not change with coupling.

Another option is to posit the presence of twisted-sector terms, i.e. that the boundary state encoding the Janus interface is built from building blocks \eqref{untwState} and \eqref{twState} obtained from a $\mathbb{T}^4$ seed theory \cite{Belin:2021nck}. Note however that this state should not describe a ``typical" interface in the symmetric orbifold. A more typical state would be built from all distinct seed states (rather than $N$ copies of $\ket{b_{\text{ND}}}^{\otimes 4}$), since the seed theory is irrational and thus itself has an infinite number of boundary states. Furthermore, the boundary entropy of a typical state would (at the orbifold point) have a divergence proportional to $N \log N$ as $N \to \infty$ stemming from both the overall $\frac{1}{\sqrt{N!}}$ normalization and the combinatorics. Thus, such boundary states would not describe an interface with a good geometric description.

From the discussion of \cite{Belin:2021nck}, the boundary state encoding the Janus interface would be ``atypical." The coefficients of the twisted-sector terms would be given by characters of some representation of $S_N$ because all $N$ seed states are identical. For the large-$N$ boundary entropy at the orbifold point to be consistent with the gravitational calculation \eqref{bdryEntropyRT} (or more specifically, its order-$N$ scaling as $N \to \infty$), the representation of $S_N$ from which the coefficients are determined would need to have a dimension
\begin{equation}
d_{\text{rep}} \sim N^{N/2},\ \ N \to \infty,
\end{equation}
thereby eliminating any imprint of the twisted-sector terms.

It would be interesting to understand more precisely the form of the boundary state describing the Janus interface. We do not anticipate the simple transport coefficients here being helpful towards this goal due to their expected sensitivity to coupling and insensitivty to combinatorics. However, we expect that calculations of boundary entropy for more stringy states (i.e. with a weak marginal coupling turned on \cite{Gaberdiel:2015uca}) could probe more of the parameter space in Figure \ref{figs:janusICFTparams}, thereby revealing more information about the twisted-sector coefficients in the boundary state. Along these lines, it would be interesting to consider the feasibility of applying the tensionless string program (e.g. \cite{Gaberdiel:2021kkp}) in a Janus background.

\subsection{Other future directions}

We also briefly describe some other directions for follow-up work.

\paragraph{Other boundary data} One could study other types of data besides entropy or transport. For example, there is ``defect complexity" \cite{Chapman:2018bqj}. This has been studied in the Janus solution through different prescriptions by \cite{Auzzi:2021nrj,Auzzi:2021ozb}. It would be interesting to see if similar quantities could be realized directly in the symmetric orbifold theory at the orbifold point.

Another approach along these lines would be to study scattering processes involving states created by extended symmetry generators (i.e. fractional Virasoro generators), which are known to probe CFT data distinct from the transport coefficients discussed here \cite{Meineri:2019ycm}.

\paragraph{Entanglement in ICFT} Transport is only facet of physics made manifest by the presence of an interface. We can study other facets of ICFT, such as entanglement entropy \cite{Karch:2021qhd,Karch:2022vot}. We can ask whether entanglement entropy generically encodes more information about the conformal interface beyond the boundary entropy of the boundary state.

\section*{Acknowledgements}

We thank Alexandre Belin, Shovon Biswas, Elena C\'aceres, Andreas Karch, and Jani Kastikainen for useful discussions. We are also grateful to Constantin Bachas, Stefano Baiguera, Shira Chapman, Andreas Karch, and Giuseppe Policastro for feedback on the draft. We deeply appreciate the extensive and useful critiques of an anonymous referee.

The work of SB was supported in part by the U.S. Department of Energy under Grant DE-SC0022021 and by a grant from the Simons Foundation (Grant 651440, AK). SS is supported by National Science Foundation (NSF) Grant No. PHY-2112725. SB and SS are also both supported by NSF Grant No. PHY-1914679.

\vfill
\pagebreak

\bibliographystyle{jhep}
\bibliography{refs.bib}
\end{document}